\documentclass[preprint,showpacs,preprintnumbers,amsmath,amssymb,superscriptaddress,floatfix]{revtex4}

\usepackage{graphicx}
\usepackage{dcolumn}
\usepackage{bm}

\begin{document}

\preprint{}

\begin{flushright}
       LYCEN 2005-05 
\end{flushright}

\title{Final results of the EDELWEISS-I dark matter search\\
with cryogenic heat-and-ionization Ge detectors}

\author{V.~Sanglard}
\email[Email address: ]{sanglard@ipnl.in2p3.fr}
\affiliation{Institut de Physique Nucl\'eaire de Lyon-UCBL,
IN2P3-CNRS, 4 rue Enrico Fermi, 69622 Villeurbanne Cedex, France}
\author{A.~Benoit}
\affiliation{Centre de Recherche sur les Tr\`es Basses Temp\'eratures,
SPM-CNRS, BP 166, 38042 Grenoble, France}
\author{L.~Berg\'e}
\affiliation{Centre de Spectroscopie Nucl\'eaire et de Spectroscopie
de Masse, IN2P3-CNRS, Universit\'e Paris XI, b\^at 108, 91405 Orsay, France}
\author{J.~Bl\"umer}
\affiliation{Institut f\"ur Experimentelle Kernphysik, Universit\"at
Karlsruhe (TH), Gaedestr. 1, 76128 Karlsruhe, Germany}
\affiliation{Forschungszentrum Karlsruhe, Institut f\"ur
Kernphysik, Postfach 3640, 76021 Karlsruhe, Germany}
\author{A.~Broniatowski}
\affiliation{Centre de Spectroscopie Nucl\'eaire et de Spectroscopie
de Masse, IN2P3-CNRS, Universit\'e Paris XI, b\^at 108, 91405 Orsay, France}
\author{B.~Censier}
\affiliation{Centre de Spectroscopie Nucl\'eaire et de Spectroscopie
de Masse, IN2P3-CNRS, Universit\'e Paris XI, b\^at 108, 91405 Orsay, France}
\author{L.~Chabert}
\affiliation{Institut de Physique Nucl\'eaire de Lyon-UCBL,
IN2P3-CNRS, 4 rue Enrico Fermi, 69622 Villeurbanne Cedex, France}
\author{M.~Chapellier}
\affiliation{CEA, Centre d'Etudes Nucl\'eaires de Saclay,
DSM/DRECAM, 91191 Gif-sur-Yvette Cedex, France}
\author{G.~Chardin}
\affiliation{CEA, Centre d'Etudes Nucl\'eaires de Saclay,
DSM/DAPNIA, 91191 Gif-sur-Yvette Cedex, France}
\author{P.~Charvin}
\affiliation{CEA, Centre d'Etudes Nucl\'eaires de Saclay,
DSM/DAPNIA, 91191 Gif-sur-Yvette Cedex, France}
\affiliation{Laboratoire Souterrain de Modane, CEA-CNRS,
90 rue Polset, 73500 Modane, France}
\author{S.~Collin}
\affiliation{Centre de Spectroscopie Nucl\'eaire et de Spectroscopie
de Masse, IN2P3-CNRS, Universit\'e Paris XI, b\^at 108, 91405 Orsay, France}
\author{M.~De~J\'esus}
\affiliation{Institut de Physique Nucl\'eaire de Lyon-UCBL,
IN2P3-CNRS, 4 rue Enrico Fermi, 69622 Villeurbanne Cedex, France}
\author{H.~Deschamps}
\affiliation{CEA, Centre d'Etudes Nucl\'eaires de Saclay,
DSM/DAPNIA, 91191 Gif-sur-Yvette Cedex, France}
\author{P.~Di~Stefano}
\affiliation{Institut de Physique Nucl\'eaire de Lyon-UCBL,
IN2P3-CNRS, 4 rue Enrico Fermi, 69622 Villeurbanne Cedex, France}
\author{Y.~Dolgorouky}
\affiliation{Centre de Spectroscopie Nucl\'eaire et de Spectroscopie
de Masse, IN2P3-CNRS, Universit\'e Paris XI, b\^at 108, 91405 Orsay, France}
\author{D.~Drain}
\affiliation{Institut de Physique Nucl\'eaire de Lyon-UCBL,
IN2P3-CNRS, 4 rue Enrico Fermi, 69622 Villeurbanne Cedex, France}
\author{L.~Dumoulin}
\affiliation{Centre de Spectroscopie Nucl\'eaire et de Spectroscopie
de Masse, IN2P3-CNRS, Universit\'e Paris XI, b\^at 108, 91405 Orsay, France}
\author{K.~Eitel}
\affiliation{Forschungszentrum Karlsruhe, Institut f\"ur
Kernphysik, Postfach 3640, 76021 Karlsruhe, Germany}
\author{M.~Fesquet}
\affiliation{CEA, Centre d'Etudes Nucl\'eaires de Saclay,
DSM/DAPNIA, 91191 Gif-sur-Yvette Cedex, France}
\author{S.~Fiorucci}
\affiliation{CEA, Centre d'Etudes Nucl\'eaires de Saclay,
DSM/DAPNIA, 91191 Gif-sur-Yvette Cedex, France}
\author{J.~Gascon}
\affiliation{Institut de Physique Nucl\'eaire de Lyon-UCBL,
IN2P3-CNRS, 4 rue Enrico Fermi, 69622 Villeurbanne Cedex, France}
\author{E.~Gerlic}
\affiliation{Institut de Physique Nucl\'eaire de Lyon-UCBL,
IN2P3-CNRS, 4 rue Enrico Fermi, 69622 Villeurbanne Cedex, France}
\author{G. Gerbier}
\affiliation{CEA, Centre d'Etudes Nucl\'eaires de Saclay,
DSM/DAPNIA, 91191 Gif-sur-Yvette Cedex, France}
\affiliation{Laboratoire Souterrain de Modane, CEA-CNRS,
90 rue Polset, 73500 Modane, France}
\author{C.~Goldbach}
\affiliation{Institut d'Astrophysique de Paris, INSU-CNRS,
98 bis Bd Arago, 75014 Paris, France}
\author{M.~Goyot}
\affiliation{Institut de Physique Nucl\'eaire de Lyon-UCBL,
IN2P3-CNRS, 4 rue Enrico Fermi, 69622 Villeurbanne Cedex, France}
\author{M.~Gros}
\affiliation{CEA, Centre d'Etudes Nucl\'eaires de Saclay,
DSM/DAPNIA, 91191 Gif-sur-Yvette Cedex, France}
\author{S.~Herv\'e}
\affiliation{CEA, Centre d'Etudes Nucl\'eaires de Saclay,
DSM/DAPNIA, 91191 Gif-sur-Yvette Cedex, France}
\author{M.~Horn}
\affiliation{Forschungszentrum Karlsruhe, Institut f\"ur
Kernphysik, Postfach 3640, 76021 Karlsruhe, Germany}
\author{A.~Juillard}
\affiliation{Centre de Spectroscopie Nucl\'eaire et de Spectroscopie
de Masse, IN2P3-CNRS, Universit\'e Paris XI, b\^at 108, 91405 Orsay, France}
\author{M.~Karolak}
\affiliation{CEA, Centre d'Etudes Nucl\'eaires de Saclay,
DSM/DAPNIA, 91191 Gif-sur-Yvette Cedex, France}
\author{C.~Kikuchi}
\affiliation{Centre de Spectroscopie Nucl\'eaire et de Spectroscopie
de Masse, IN2P3-CNRS, Universit\'e Paris XI, b\^at 108, 91405 Orsay, France}
\author{A.~de~Lesquen}
\affiliation{CEA, Centre d'Etudes Nucl\'eaires de Saclay,
DSM/DAPNIA, 91191 Gif-sur-Yvette Cedex, France}
\author{M.~Luca}
\affiliation{Institut de Physique Nucl\'eaire de Lyon-UCBL,
IN2P3-CNRS, 4 rue Enrico Fermi, 69622 Villeurbanne Cedex, France}
\author{J.~Mallet}
\affiliation{CEA, Centre d'Etudes Nucl\'eaires de Saclay,
DSM/DAPNIA, 91191 Gif-sur-Yvette Cedex, France}
\author{S.~Marnieros}
\affiliation{Centre de Spectroscopie Nucl\'eaire et de Spectroscopie
de Masse, IN2P3-CNRS, Universit\'e Paris XI, b\^at 108, 91405 Orsay, France}
\author{L.~Mosca}
\affiliation{CEA, Centre d'Etudes Nucl\'eaires de Saclay,
DSM/DAPNIA, 91191 Gif-sur-Yvette Cedex, France}
\author{X.-F.~Navick}
\affiliation{CEA, Centre d'Etudes Nucl\'eaires de Saclay,
DSM/DAPNIA, 91191 Gif-sur-Yvette Cedex, France}
\author{G.~Nollez}
\affiliation{Institut d'Astrophysique de Paris, INSU-CNRS,
98 bis Bd Arago, 75014 Paris, France}
\author{P.~Pari}
\affiliation{CEA, Centre d'Etudes Nucl\'eaires de Saclay,
DSM/DRECAM, 91191 Gif-sur-Yvette Cedex, France}
\author{L.~Schoeffel}
\affiliation{CEA, Centre d'Etudes Nucl\'eaires de Saclay,
DSM/DAPNIA, 91191 Gif-sur-Yvette Cedex, France}
\author{M.~Stern}
\affiliation{Institut de Physique Nucl\'eaire de Lyon-UCBL,
IN2P3-CNRS, 4 rue Enrico Fermi, 69622 Villeurbanne Cedex, France}
\author{L.~Vagneron}
\affiliation{Institut de Physique Nucl\'eaire de Lyon-UCBL,
IN2P3-CNRS, 4 rue Enrico Fermi, 69622 Villeurbanne Cedex, France}
\author{V.~Villar}
\affiliation{CEA, Centre d'Etudes Nucl\'eaires de Saclay,
DSM/DAPNIA, 91191 Gif-sur-Yvette Cedex, France}

\collaboration{EDELWEISS Collaboration}

\date{\today}

\begin{abstract}
The final results of the EDELWEISS-I dark matter
search using cryogenic heat-and-ionization Ge detectors are presented.
The final data sample corresponds to an increase by a factor five
in exposure relative to the previously published results. A recoil 
energy threshold of 13~keV or better was achieved with three 320~g 
detectors working simultaneously over four months of stable 
operation. Limits on the spin-independent cross-section for the 
scattering of a WIMP on a nucleon are derived from an accumulated 
fiducial exposure of 62~kg$\cdot$d. 
\end{abstract}

\maketitle

\section{\label{sec:introduction}Introduction}
The search for the particles constituting the non-baryonic
dark matter content of our Universe is a domain of intense experimental
activities (see e.g. Ref.~\cite{bib-morales} for a review). In the so-called
direct search~\cite{bib-lewinsmith}, one looks for nuclear recoils induced by the scattering
on terrestrial targets of Weakly Interacting Massive Particles (WIMPs)
that are part of the dark matter halo of our Galaxy.
The Minimal SuperSymmetric Model (MSSM), where the WIMP is the 
neutralino~\cite{bib-jungman}
(Lightest Supersymmetric Particle), predicts scattering rates ranging 
from one interaction per kilogram of detector per week, to less than one interaction 
per ton per year~\cite{bib-susy}. The experimental challenge is to discriminate
these rare events from the much larger backgrounds from natural radioactivity.
The expected recoil energies range typically from a few keV to a few tens 
of keV, a relatively
low energy scale for usual particle physics detectors. Up to now,
the best sensitivities have been obtained by cryogenic detectors with
nuclear recoil identification capabilities~\cite{bib-edw2000,bib-edw2002,bib-cdms-suf,bib-cdms-soudan,bib-cresst}.
In these techniques a heat (or phonon) channel measures the energy
deposit independently of the nature of the recoiling particle. A second
channel measures the ionization yield in a semiconductor crystal (CDMS~\cite{bib-cdms-suf,bib-cdms-soudan}
and EDELWEISS~\cite{bib-edw2000,bib-edw2002}) or the light yield of a scintillating
crystal (CRESST~\cite{bib-cresst}). The overwhelming background from
$\gamma$ and $\beta$ radiation is reduced by factors larger than
1000 by exploiting the fact that electron recoils have larger ionization
or scintillation yields than nuclear recoils.\\
The previous EDELWEISS~\cite{bib-edw2002} results
were the first to probe the predictions of a first set of supersymmetric
models. Since then, CDMS~\cite{bib-cdms-soudan} has published new results
with a factor four improvement in sensitivity. Limits obtained by 
the CRESST experiment using W recoils~\cite{bib-cresst} show a sensitivity 
similar to that of EDELWEISS.\\
The published results of EDELWEISS were obtained
using single 320~g heat-and-ionization Ge detectors, with accumulated
fiducial exposures of 5.0~\cite{bib-edw2000} and 13.6~kg$\cdot$d~\cite{bib-edw2002}.
Since then the experiment has completed its phase I, reaching its
goal to operating simultaneously three detectors in low-background run
conditions over a period of several months. In this paper
we present the final results of the EDELWEISS-I experiment. A new
sample representing a fiducial exposure of 48.4~kg$\cdot$d is added
to the 13.6~kg$\cdot$d of data presented in Refs.~\cite{bib-edw2000,bib-edw2002}.
New limits are established with the total fiducial exposure of 62~kg$\cdot$d,
superseding the previously published results~\cite{bib-edw2002}. The
origins of the possible backgrounds limiting the sensitivity of the
present setup are discussed. Two key achievements are pursued. The first 
is to reach
an energy threshold better than 20~keV for the detection and discrimination
of nuclear recoils. The second is the identification of the nature
of possible backgrounds that could appear in the sensitivity domains
beyond those first explored in Ref.~\cite{bib-edw2002}. In addition to the presentation 
of the sensitivity reached by the EDELWEISS-I experiment, this work is also an essential
preparation for the more ambitious phase II, where up to 120 detectors
will be operated in a larger cryostat in an optimized low-radioactivity
environment.
\section{\label{sec:setup}Experimental setup}
The experimental setup is described in detail in Refs.~\cite{bib-edw2000,bib-edw2002,bib-edw-calib}.
Only the most relevant aspects as well as the improvements made since
then will be summarized in this section.
\subsection{\label{sec:shielding}Shielding}
The experiment is located in the Laboratoire Souterrain
de Modane (LSM) in the Fr\'ejus tunnel under the French-Italian Alps.
The rock coverage, equivalent to 4800~m of water, reduces the cosmic
muon flux to 4.5 muons per day for a horizontal detector surface of 
1~m$^{2}$. The neutron flux $\Phi_n$ in the 2-10~MeV range
is $\Phi_n\sim$~1.6$\times10^{-6}$~/cm$^{2}$/s~\cite{bib-verene,bib-edwneutron}.
The detectors are protected from the surrounding $\gamma$-ray background
with 10~cm of Cu and 15~cm of Pb~\cite{bib-bellefon}. Pure nitrogen
gas is circulated within this shield to reduce radon accumulation.
An additional 7~cm thick internal roman lead shield screens the detectors
from radioactive electronic components. The entire setup is protected
from the neutron background by an outer 30~cm paraffin shield.
\subsection{\label{sec:detector}Detectors}
Three 320~g cryogenic heat-and-ionization Ge detectors
(70~mm in diameter and 20~mm in height with edges beveled at an angle of 
45$^{\circ}$) are operated simultaneously
in a low-background dilution cryostat~\cite{bib-bellefon} running at
a regulated temperature of 17.00~$\pm$~0.01~mK. They are individually
housed in separate 1~mm thick Cu casings, the distance between the
Ge surfaces being 13~mm. For the heat measurement, a NTD-Ge thermometric
sensor is glued on each detector. For the ionization measurement,
the detectors are equipped with two Al electrodes. One is segmented
to define two regions, a central part and a guard ring~\cite{bib-edw-calib}.
The applied collection voltage $V_{bias}$ is either $+$6.34~V or $-$4.0~V.\\
Over the years, seven detectors have been used. Their characteristics 
are listed in Table~\ref{tab:tab-Detectors}.
For the first three, labeled GeAl6, GeAl9 and GeAl10, the Al electrodes 
are directly sputtered on the Ge crystal. As shown in Ref.~\cite{bib-shutt},
and also observed in Ref.~\cite{bib-edw2002}, a better charge collection
is achieved with a Ge or Si amorphous layer under the electrodes.
Therefore, only one of the GeAl detectors was used in a short low-background
run~\cite{bib-edw2000}. The two detectors with Ge amorphous layers are
labeled GGA1 and GGA3, and the two detectors with Si amorphous layers,
GSA1 and GSA3.\\ 
We present here the results of two new runs in addition
to the two low-background runs for which results have been published
in Refs.~\cite{bib-edw2000,bib-edw2002}. These two new runs, named 2003i
and 2003p (see Sec.~\ref{sec:acquisition}), have been recorded with a stack comprising the three detectors
GSA3, GSA1 and GGA3. The experimental configurations are described
in the following.
\subsection{\label{sec:acquisition}Data acquisition}
The numerical acquisition system is based on a commercial
PXI system. For each of the three detectors the measured quantities
are two ionization signals from the center and guard electrodes, and
one heat signal from the NTD-Ge thermometric sensor. The analog signals
are pre-amplified at a cold-FET stage, amplified at ambient temperature,
filtered to avoid aliasing and then digitized on two PXI cards. The
six ionization channels are recorded with a multiplexed 16-channel
PXI-6070E card with 12-bit precision at a sampling rate of 200~ksample/s.
The heat signals - for which the time constants are slower by three
orders of magnitude relative to the ionization signals - are recorded at a
rate of 1 or 2~ksample/s, depending on the data takings, with a slower 
PXI-6052E card with 16-bit precision.\\
The data are then transferred via a dedicated 1.5~Gbit/s
optical link to the bi-Xeon computer running the acquisition program.
To optimize the signal-to-noise ratio, the digitized signal of each
channel first passes through a specific IIR (Infinite Impulse Response)
numerical bandpass filter tailored to the main features of the noise
spectrum. The trigger is defined numerically by requesting that the
absolute value of any channel exceeds a given threshold. Up to 2002
and in the first run in 2003 (run 2003i), the trigger was based on the ionization
channels. In the last run in 2003 (run 2003p), it was based on the phonon 
(or heat) channel.
This phonon trigger configuration results in a better sensitivity
at low energy.\\
The ionization trigger setup has already been described
in previous publications~\cite{bib-edw2000,bib-edw2002}. It basically scans 
data blocks in a circular buffer. If the trigger conditions are fulfilled
for any ionization channels, the relevant data are saved to disk.\\
The phonon trigger setup first requires that one
of the three heat channels exceeds a predefined level. When a trigger is
found the relevant ionization information lies in the past, due to
its $\sim$~1000 times faster rise-time. Hence the two corresponding
center- and guard-channel buffers are scanned back 20~ms to find
the most appropriate signal candidate. This is achieved by performing
a convolution of the data with a template of an ionization signal
built offline. The maximum of convolution gives the position of the
ionization signal. The size of the "search zone"
of 20~ms corresponds to the total heat signal rise-time with a safety
margin of $\sim$~30~\%.\\
When the position of the ionization signal is localized,
the relevant portion of data for all channels on each detector is
saved to disk, as well as the value of the time difference between
the ionization and heat channels computed online. The stored samples
correspond to 10~ms of ionization data and 1~s of heat data. 
In addition, for
each event its absolute time of occurrence, the instantaneous
temperature of the dilution refrigerator, the results of the online
convolution performed by the trigger system, and a bit pattern corresponding
to the detectors with a heat signal above the trigger threshold in
a 120~ms window are recorded. To avoid triggering twice on the same event, the
minimum time between two events is set to 0.76~s, resulting in an equivalent 
dead time per recorded event.\\
The data acquisition is automatically stopped for
12~min every 3~hours. This corresponds to the time where the electrodes
are short-circuited in order to prevent the accumulation of space
charge~\cite{bib-regeneration}.
\subsection{\label{sec:processing}Signal Processing}
\begin{figure}[tbp]
\includegraphics[scale=0.85]{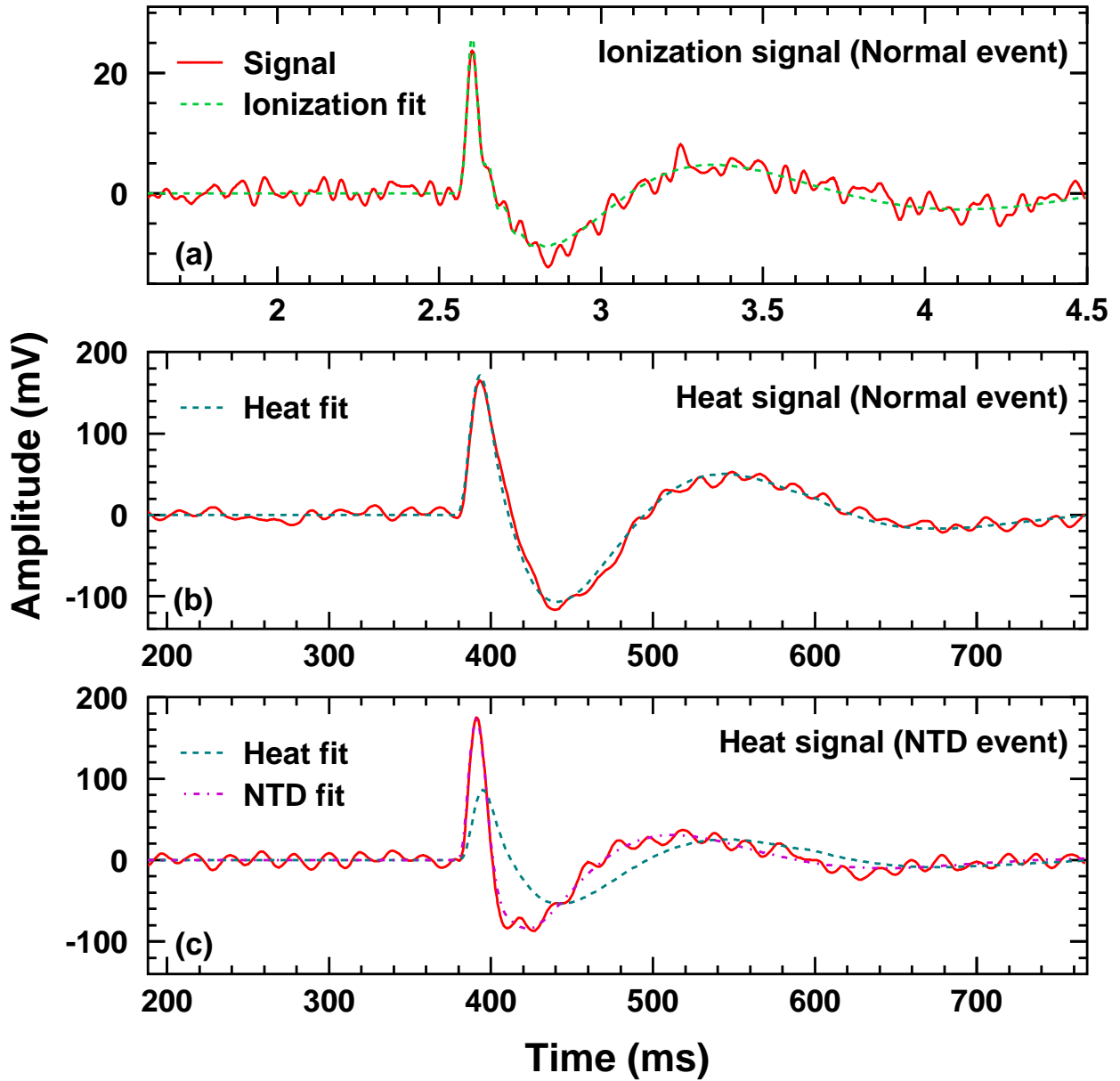}
\caption{\label{fig:norm-ntd}Example of filtered heat and ionization pulses for $\sim$~10~keV$_{ee}$ signals (full lines) together
with the template fit (dashed lines) for an ionization (center electrode) signal (a) and for the corresponding heat signal (b). 
In (c) is shown an example of a NTD event (see text) together with the template fits for a normal 
heat signal (dashed line) and for a NTD signal (dash-dotted line)}
\end{figure}

The stored events are re-processed offline. A detailed
description of the signal processing can be found in Ref.~\cite{bib-edw-calib}.
In the offline analysis, templates of ionization and heat signals
are adjusted with the constraint of the simultaneity of the center,
guard and heat signals in a given detector. The piled-up pulses, practically 
negligible in low-background data, are more numerous in calibration runs and 
are taken into account with the simultaneous
adjustment of more than one template to each event. The templates
are built with a sample of selected 122~keV events from a $^{57}$Co
source, one template for each channel and each detector. It was verified that the templates 
did not vary with time, except at the beginning of the run 2003p when the digital filters 
have been modified (see Sec.~\ref{sec:data-new}).
On Fig.~\ref{fig:norm-ntd}(a) and (b) are shown examples of filtered 
ionization and heat pulses (full lines), respectively, for
$\sim$~10~keV$_{ee}$ signals in the detector GGA3, together with the template
fits (dashed lines). These low energy signals are well 
modelled by the 122~keV pulse template.
The $\chi^2$ of the fits do not depend on the pulse amplitude, showing that the
pulse shape does not vary with amplitude, at least up to $\sim$~300~keV.
The cross-talk between the two electrode signals is less than
4~\% and remains constant through time. It is treated as described
in Ref.~\cite{bib-edw-calib}.\\
In the phonon trigger data, some events are due
to the internal radioactivity of the NTD sensor. For these events,
the deposited energy in the NTD is not accompanied by an ionization
signal. These so-called NTD events occur at a rate of $\sim$~0.5~mHz.
In this case, the shape of the heat signal is different. To identify
these events, each heat sample is processed twice: first with a normal
template, and then with the template of a NTD-event pulse. 
This NTD pulse template is built using a small sample of such events 
with a large amplitude, detected by the absence of ionization signal and 
large $\chi^2$ values for the fit with the normal template. Fig~\ref{fig:norm-ntd}(c) 
shows an example of such NTD pulse (full line) together with the normal template 
fit (dotted line) and the NTD template fit (dash-dotted line).
Most NTD
events are rejected, with no loss of efficiency ($<$~0.1~\%) down to a 
recoil energy of 10~keV,
with a test on the ratio of the $\chi^{2}$ of the two fits. The remaining
NTD events are removed by an offline cut on the ionization energy,
included in the determination of the efficiency discussed below.
\section{\label{sec:calibration}Detector Calibration}
\begin{figure}[tbp]
\includegraphics[scale=0.85]{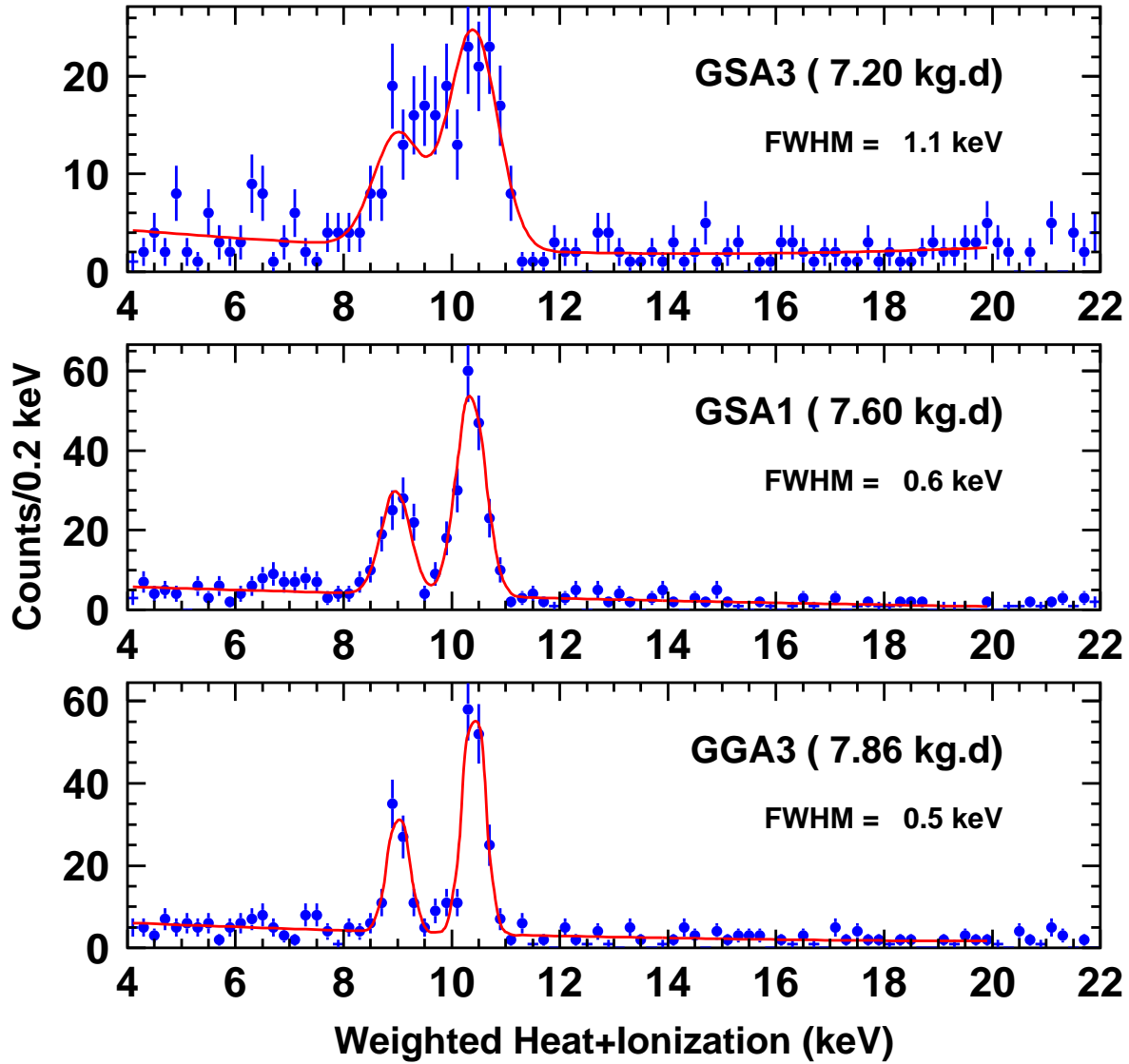}
\caption{\label{fig:fig-10kev}Low-energy part of the spectrum
recorded in the fiducial volume of the three detectors. 
The energy is calculated as the sum of the ionization and heat
channels, weighted by their resolution squared. The peaks at 8.98
and 10.37~keV correspond to the de-excitation of the cosmogenic activation
of $^{65}$Zn and $^{68}$Ge in the detectors, and the $^{71}$Ge
activation that follows neutron calibrations. The lines correspond to a Gaussian 
fit with the indicated value of FWHM resolutions.}
\end{figure}

\begin{figure}[tbp]
\includegraphics[scale=0.85]{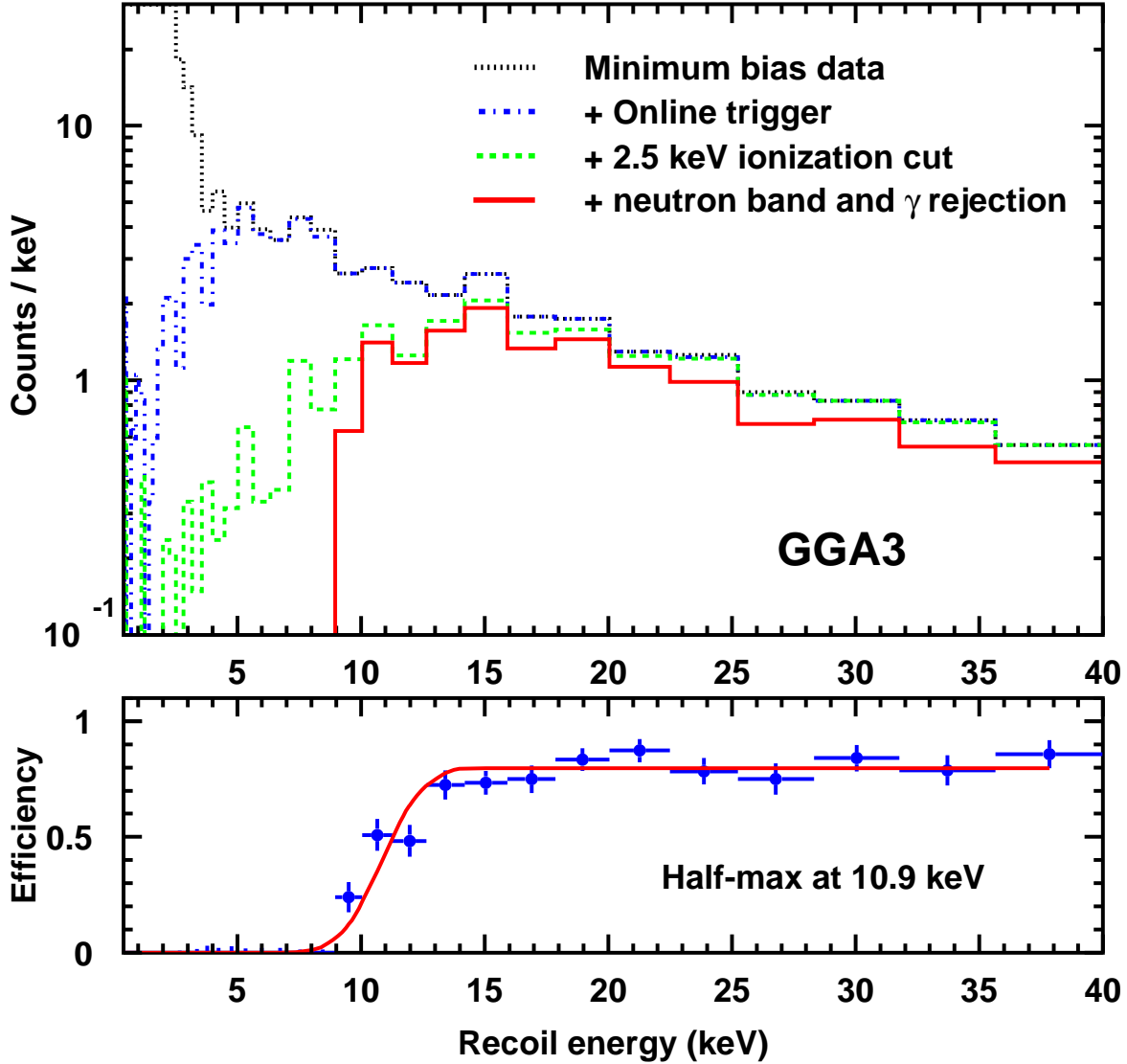}
\caption{\label{fig:fig-effic}Measurement of the efficiency
as a function of recoil energy for the detector GGA3 in the run 2003p
configuration, using neutron-coincidence data from a $^{252}$Cf calibration.
Top: spectrum as a function of energy with different cuts.
Dotted: minimum bias (selection based only on the presence of a neutron
in the other two detectors); dot-dashed: adding the condition that
the heat signal is above threshold; dashed: adding the 2.5~keV$_{ee}$ ionization
cut; full line: adding the $\pm$~1.65$\sigma$ and $<$~$-$3.29$\sigma$
nuclear and electron recoil requirements (see text). Bottom: resulting
efficiency as a function of energy. The maximum value is not 90~\%
because the data are not corrected for the effect of neutron-$\gamma$
coincidences.}
\end{figure}

As described in Ref.~\cite{bib-edw-calib}, the heat
signal $E_{H}$ is calibrated in keV-electron-equivalent (keV$_{ee}$).
The center and guard electrode signals are also calibrated in keV$_{ee}$
and added to give the total ionization amplitude $E_{I}$. From these
two measurements, the recoil energy $E_{R}$ and the ionization quenching
$Q$ are deduced event-by-event by correcting for the Joule heating
due to the applied voltage $V_{bias}$~\cite{bib-luke}:
\begin{equation}
E_{R}=(1+\frac{{|V_{bias}|}}{\epsilon_{\gamma}})E_{H}-\frac{{|V_{bias}|}}{\epsilon_{\gamma}}E_{I}
\label{eq:Er}
\end{equation}
\begin{equation}
Q=\frac{E_I}{E_R}
\label{eq:Q}
\end{equation}
where $\epsilon_{\gamma}\cong$~3.0~V for Ge, the applied voltages
are $V_{bias}$~=~$+$6.34~V for GeAl6 and $V_{bias}$~=~$-$4.0~V for the GGA and GSA detectors.\\
The detector calibration follows the method described
in Refs.~\cite{bib-edw2000,bib-edw2002,bib-edw-calib}. The calibration of ionization
signals at 122~keV is performed using a $^{57}$Co source. It is
checked with a $^{137}$Cs source and the X-ray data described later
that the ionization channel is linear from 9 to 662~keV. 
The calibration runs were performed on a monthly basis.
Over the
entire running period, there is no observable drift in time of the
ionization gains. After a first calibration at 122~keV, the heat
signal is calibrated by imposing that its ratio to the ionization
signal should be unity for all $\gamma$-rays. Non-linearities on the 
heat channel between
0 and 662~keV are determined using $^{137}$Cs data. The dependence
in time of the overall heat gain is obtained by monitoring closely
the ratio of the ionization and heat signals as a function of time.
The largest drifts in heat gain are less than a few percent per hour,
and either occur on occasional failure of the temperature regulation
system, or within five hours after refilling the cryostat with liquid
He. Drifts of up to 10~\% in heat gain are corrected as linear
functions of time in order to avoid abrupt changes of calibration
constants during runs. Data sets with larger drifts are discarded.\\
Thanks to the improved heat resolution and the increased
statistics, it was possible to study in more detail than in the 
previous work the calibration at very low energy. Summed X-ray peaks
are emitted following the electron capture decay of $^{65}$Zn 
(T$_{1/2}$~=~244~d)
and $^{68}$Ge (T$_{1/2}$~=~271~d) due to the activation
of the detectors by cosmic rays before their installation at the LSM,
and of $^{71}$Ge (T$_{1/2}$~=~11.4~d) activated following
$^{252}$Cf neutron source calibrations. These total K-shell energy
peaks at 8.98 and 10.37~keV for Zn and Ge decays, respectively, are
clearly seen in Fig.~\ref{fig:fig-10kev}. They are particularly useful
to verify the accuracy of the energy calibration at low energies. 
For example, using the $^{57}$Co calibration at 122 and 136~keV, the 
energies of the 8.98 and 10.37~keV X-ray peaks are reproduced within 
0.1~$\pm$~0.1~keV.\\
To select events occurring in the central part of
the detector, where the electric field is the most uniform and the
detector better shielded from its environment, a fiducial cut is made
by requiring that at least 75~\% of the charge signal comes from the center
electrode. As measured in Ref.~\cite{bib-edw-calib}, using data recorded
with a $^{252}$Cf neutron source, this requirement corresponds to
55~$\pm$~1~\% of the total volume for GeAl6 and 58~$\pm$~1~\% of the total
volume for the GGA and GSA detectors, with slightly different electrode 
designs. Conservatively, the adopted
values are 54~\% and 57~\%.\\
Calibrations with a $^{252}$Cf neutron source have
been performed for each detector and each run configuration in order
to establish the zone in the $Q$ vs $E_{R}$ plane corresponding
to a 90~\% efficiency to detect nuclear recoils induced by neutron
scattering. It has been verified that, in all cases, the nuclear recoil
band is well described with the parametrization used 
previously~\cite{bib-edw2000,bib-edw2002,bib-distefano}.
Namely, the distribution is Gaussian, centered around $Q=0.16E_{R}^{0.18}$,
and its width $\sigma_{Q}$ is given by the propagation of the experimental
resolutions on $E_{H}$ and $E_{I}$ (see Table~\ref{tab:tab-Data-resos}),
smeared by an additional spread
$C$, see Eq.~11 from~\cite{bib-edw-calib}. The constant $C$ represents
the effects of multiple neutron scattering and energy straggling in
the stopping of the Ge recoils. The experimental $\sigma_{Q}$ in
neutron calibrations are well reproduced with $C$=~0.035. The width
of the band for WIMP-induced recoils should not be altered by multiple
scattering, but in Ref.~\cite{bib-edw-calib} it was shown that the band
measured in neutron calibration is a conservative estimate of the
90~\% efficiency region for WIMP induced recoils.\\
The same neutron calibrations yield precise measurements
of the nuclear recoil detection efficiency as a function of $E_{R}$
(and in particular, close to threshold). Here, the
large number of neutron-neutron coincidences is used as a source of "minimum
bias events". In practice, for a given detector this sample is defined
as events where a neutron has been recorded in one of the other two
detectors. In previous works~\cite{bib-edw2000,bib-edw2002,bib-edw-calib}, samples
defined in this way were used to measure the efficiency as a function
of the signal used for triggering, $E_{I}$. Here, 
it is measured as a function of recoil energy $E_R$. In order to eliminate NTD 
events and accept only events with charge amplitude above noise, a 
2.5~keV$_{ee}$ cut is applied on the total ionization amplitude of 
each detector. The trigger efficiency measured after the online phonon 
trigger and the minimum ionization cut is illustrated for detector GGA3 
in Fig.~\ref{fig:fig-effic}.\\ 
The top panel of this figure 
shows the $E_{R}$ distribution in GGA3 for minimum bias events (dotted
histogram), as well as for events where the online trigger has detected
a heat signal in GGA3 (dot-dashed histogram), and for events where
in addition the ionization signal exceeds 2.5~keV$_{ee}$ (dashed
histogram). In coincident data, the time of all ionization signals
is given by the largest amplitude charge signal in any of the detectors. It is thus
possible to identify accurately ionization signals that are below
2.5~keV$_{ee}$ and to observe (dot-dashed histogram) low-energy
events that would be otherwise lost in single detector data. Most
of the inefficiency at low energy comes from the 2.5~keV$_{ee}$
cut. Fig.~\ref{fig:fig-effic} also shows the effect of the two additional
cuts on $Q=E_{I}/E_{R}$ in the final analysis (full-line histogram): the
first one to select neutrons ($\pm$~1.65$\sigma$,corresponding to 90~\% efficiency) 
and the second one to reject $\gamma$-rays ($<$~$-$3.29$\sigma$,
corresponding to 99.9~\% rejection for a Gaussian distribution in $Q$ centered
at one). The truncation
at 9~keV is due to the $\gamma$-ray rejection cut. The cumulative
effect of the trigger and the selection cuts on the efficiency as
a function of recoil energy for nuclear recoils is shown in the lower
panel of Fig.~\ref{fig:fig-effic}, where this quantity is obtained
from the ratio of the full-line and dotted histograms in the top panel.
At the plateau, the measured
ratio is approximately 80~\%. After correcting for neutron coincidences
with $\gamma$-rays as in for example, inelastic (n, $\gamma$) scattering,
it is verified that the defined band contains 90~\% of the elastic nuclear 
recoil interactions.\\
The "threshold energy" is defined as the energy
at which the efficiency reaches half its maximum value. It is a relevant
variable for comparing detectors among themselves, and with the
detector simulations (see Sec.~\ref{sec:simu} for details). For
GGA3 in the phonon trigger configuration, the energy threshold is
11~$\pm$~1~keV (see Fig.~\ref{fig:fig-effic}). The measured values for
the detectors in the different configurations where coincident neutron
measurements were possible are listed in Table~\ref{tab:tab-Data-sets}.
In the run 2003p, the recoil energy thresholds on the three detectors 
were better
than 13~keV. This is better than what is achieved in the run 2003i,
where the corresponding values range from 14 to 23~keV. The improvements
are due to three factors. Firstly, the baseline heat resolution is generally
better than the ionization one (see Table~\ref{tab:tab-Data-resos}).
Secondly, the ionization signal for nuclear recoils is significantly
reduced by the quenching effect. Thirdly, ionization signals with a lower 
amplitude can be recorded in the phonon trigger configuration because this trigger 
requires a coincidence between a phonon trigger, with better sensitivity, 
and an ionization signal greater than 2.5~keV$_{ee}$ searched on a short (20~ms) 
time window immediately preceding the time at which the heat signal is detected. 
\section{\label{sec:Data-sets}Data sets}
\subsection{\label{sec:selection}WIMP candidate selection}
An event in a detector is considered
as a WIMP candidate in the fiducial volume if it obeys the following criteria:
\begin{enumerate}
\item More than 75~\% of the charge is collected on the
central electrode.
\item The ionization signal $E_{I}$ exceeds the ionization
threshold value (listed in Table~\ref{tab:tab-Data-sets}).
\item The $Q$ and $E_{R}$ values are inside the $\pm$~1.65$\sigma$
(90~\%) nuclear recoil band.
\item The $Q$ and $E_{R}$ values are outside the $\pm$~3.29$\sigma$
(99.9~\%) electron recoil band.
\item Only this detector participates in the trigger. 
\end{enumerate}
For each detector and run configuration, the nuclear
and electron recoil bands are calculated using the corresponding experimental
resolutions (see Table~\ref{tab:tab-Data-resos}). As stated earlier,
it was verified that the $\pm$~1.65$\sigma$ neutron band contains
90~\% of neutron scattering events in $^{252}$Cf calibrations, excluding
inelastic events where some energy is deposited by an additional $\gamma$-ray.
Given the expected statistics in the low-background runs, a safe rejection
of $\gamma$-rays requires to extend the width of the electron recoil
band beyond 2 or 3$\sigma$, depending on $E_{R}$. Although the $Q$
distributions in $\gamma$-ray calibrations appear to be Gaussian
up to 3$\sigma$, it has not been possible to accumulate enough statistics
to verify this assertion with precision. For safety, a width of $\pm$~3.29$\sigma$
is adopted, which corresponds to a 99.9~\% rejection for a Gaussian
distribution. The effective rejection may not be as good, but the
procedure yields conservative upper limits on WIMP-nucleon scattering
cross-sections (see Sec.~\ref{sec:limite}).
\subsection{\label{sec:data-previous}Previous data sets}
In 2000 and 2002, two low-background runs have been
performed. The results have been published in Refs.~\cite{bib-edw2000}
and~\cite{bib-edw2002}, respectively. These data sets have not been
reprocessed. However, the nuclear recoil selection has been modified
in order to be consistent with the one described above. The only modification
relative to Refs.~\cite{bib-edw2000,bib-edw2002} is the removal of the low-energy
bound on $E_{R}$ that was previously set to either 20 or 30~keV,
depending on the energy for which the efficiency value is approximately constant
with energy and close to 90~\%. In the present work, this lower bound is replaced by
the more natural constraint given by the 3.29$\sigma$ $\gamma$-ray
rejection and the ionization threshold. The reduced efficiency below
20 and 30~keV is taken into account in a later stage of the analysis
(see Sec.~\ref{sec:simu}).\\
The run 2000 comprises three configurations (see
Table~\ref{tab:tab-Data-sets}) with ionization thresholds of 5.7, 9.0
and 11.0~keV$_{ee}$. The corresponding fiducial exposures are respectively
3.80, 0.63 and 0.60~kg$\cdot$d, for a total of 5.03~kg$\cdot$d.
The data are shown in Fig.~2 of Ref.~\cite{bib-edw2000}. No nuclear
recoil candidates are observed above the analysis threshold of 30~keV
used in Ref.~\cite{bib-edw2000}. Two events fall within the selection
defined in Sec.~\ref{sec:selection}: at $(E_{R},Q)$~=~(22.5~keV, 0.367)
and (25.1~keV, 0.312), both recorded in the first configuration. Another 
event is observed at (29.3~keV, 0.420) in the third configuration. 
It is excluded by the 3.29$\sigma$ $\gamma$-ray
rejection corresponding to this configuration.\\
The run 2002 corresponds to a fiducial exposure
of 8.6~kg$\cdot$d with an ionization threshold of 3.5~keV$_{ee}$.
As can be seen in Fig.~3 of Ref.~\cite{bib-edw2002}, five events satisfy 
the new selection criteria, two of them having recoil energies above
15~keV (18.8 and 119~keV).
\subsection{\label{sec:data-new}New experimental conditions and data sets}
\begin{figure}[tbp]
\includegraphics[scale=0.85]{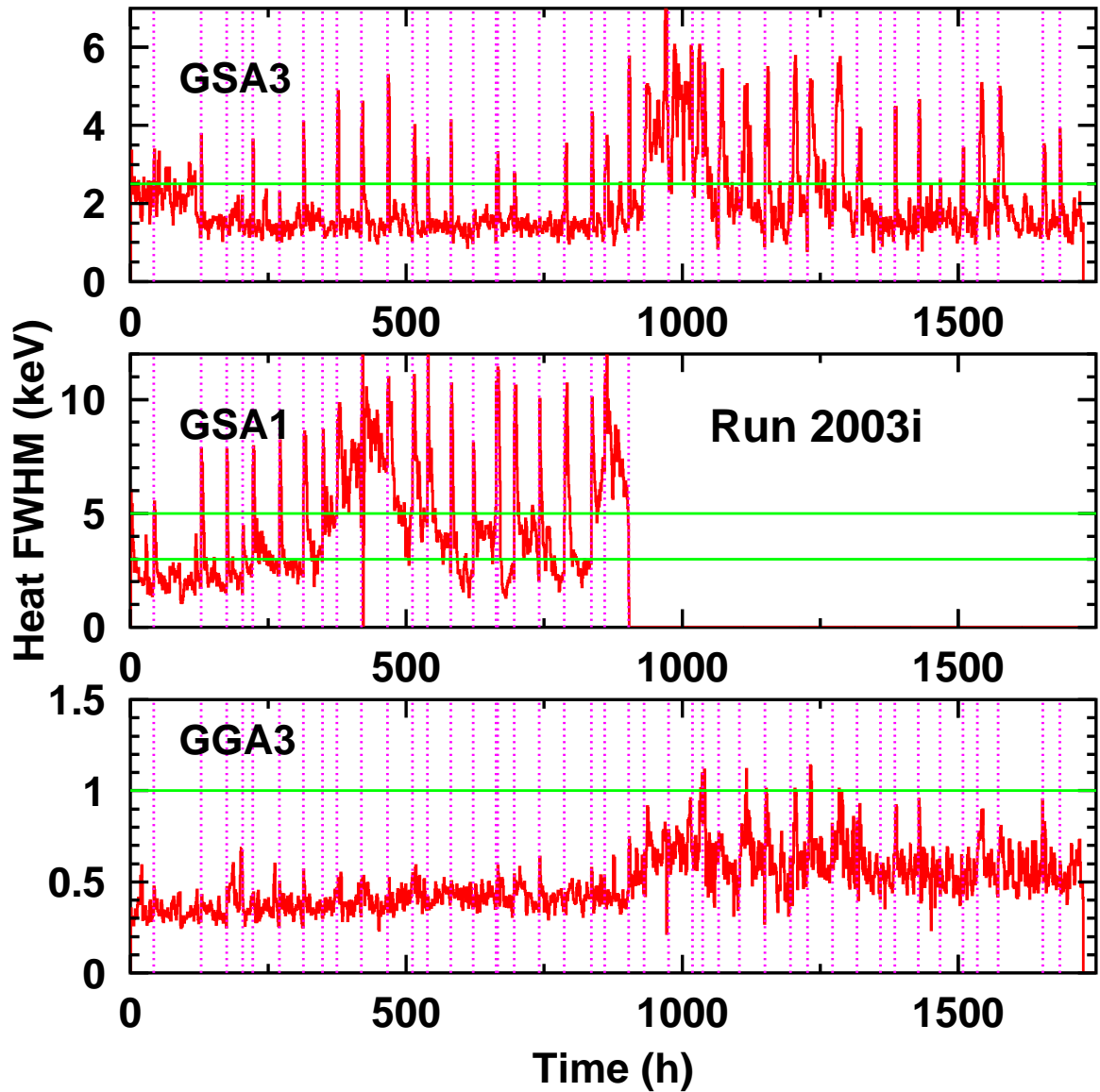}
\caption{\label{fig:fig-datquali}Baseline FWHM resolution on the heat channel
of the three detectors in the run 2003i as a function of time in hours
since the beginning of the run. The resolution is evaluated by 3-hours
intervals centered on each hour. The dotted lines represent times
when the cryostat was re-filled with liquid He. The full lines
represent FWHM cuts.}
\end{figure}

\begin{figure}[tbp]
\includegraphics[scale=0.85]{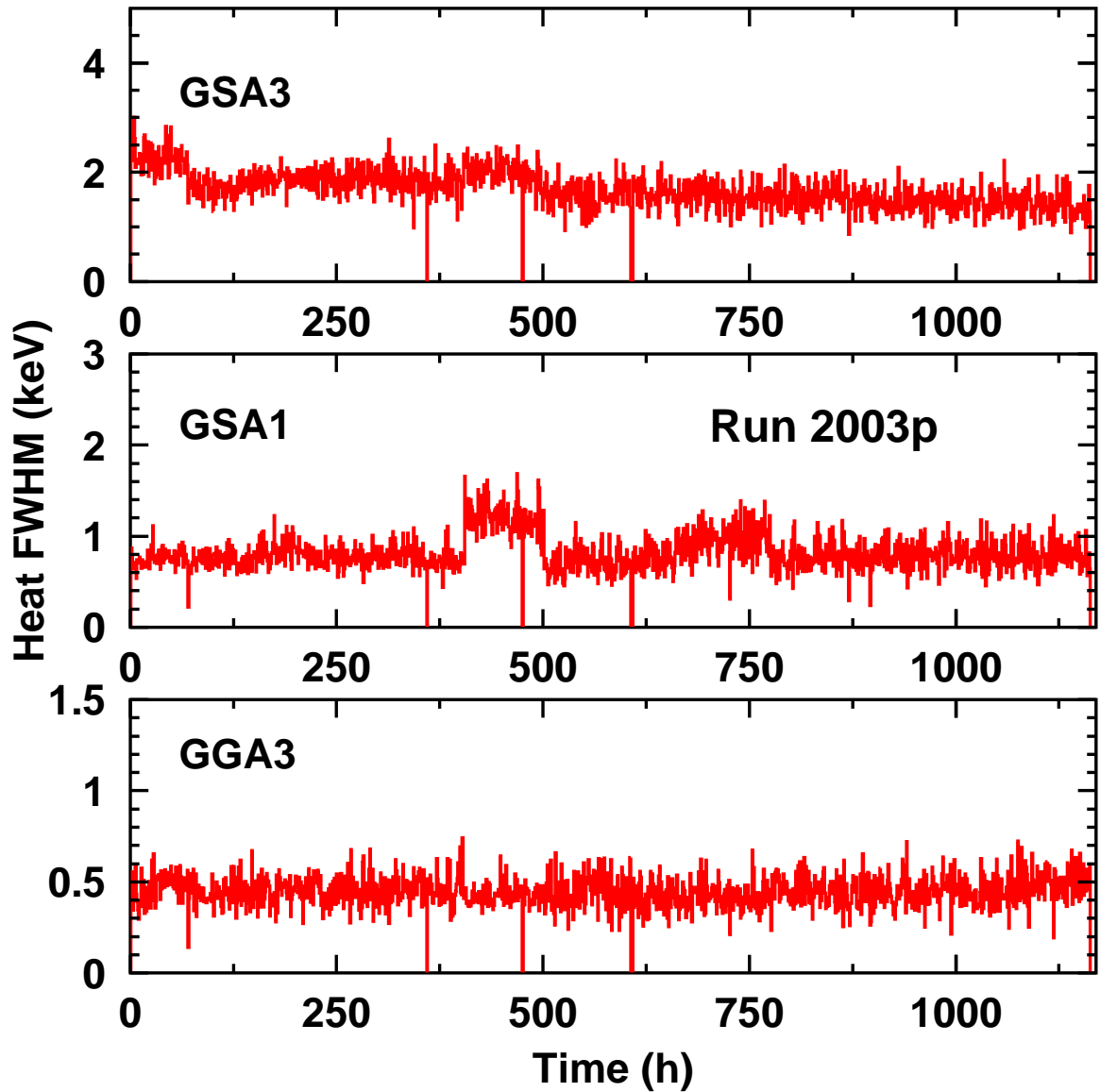}
\caption{\label{fig:fig-datqualp}Baseline FWHM resolution on the heat channel
of the three detectors in the run 2003p as a function of time in hours
since the beginning of the run. The resolution is evaluated by 3-hours
intervals centered on each hour.}
\end{figure}

For the new runs, the experimental setup was upgraded
in order to address the three following points. 
\begin{itemize}
\item{Firstly, the results of the run 2002~\cite{bib-edw2002} together with 
studies of Ref.~\cite{bib-shutt} suggested that the presence of an amorphous layer
under the metallic electrodes improved the charge collection.
This was later confirmed with short test runs of detectors with and
without an amorphous layer, either in Ge or Si. Consequently, a stack
of three detectors with amorphous layers was assembled (GSA3, GSA1
and GGA3) for an extended low-background run.} 
\item{Secondly, the runs 2000
and 2002 had been limited by cryogenic problems caused indirectly
by the regular disruptions associated with the manual procedure for
the filling of cryogenic fluids. It was therefore decided to install
an automatic liquid He filling system with an associated monitoring
system.}
\item{Finally, it had been noticed that the baseline noise levels
on the ionization and heat channels were very sensitive to the quality
of the electrical connections between the detectors, the cold FETs
and the warm amplifiers. For the new low-background runs, the wiring
was re-designed for an improved reliability.}
\end{itemize}
The new data sets are separated into two running
periods. \\
In the first one (run 2003i) the automatic filling system
was being commissioned. For safety reasons, the automatic monitoring
of the liquid He level was permanently on, at the expense of additional
noise on the signals. Despite the improvements in wiring, large fluctuations
in baseline noise were still observed on the GSA1 heat channel, sometimes
reaching levels at which the induced cross-talk observed in the other
detectors degraded significantly their resolutions. As shown later,
this is particularly true just after He refilling, indicating a high
sensitivity to microphonic noise.\\
The second running period (run 2003p) corresponds
to a new stable configuration, where the problems associated with
noise due to the He monitoring system and other sources of microphonics
were cured. As the He filling system had proved its reliability, the
monitoring was switched off during the low-background runs. The sensitivity
to the microphonic noise was reduced when the stray capacitance between
the cold FETs and the warm amplifiers was decreased by replacing a
patch panel interface between them with soldered connections. These
improvements were performed and tested in a few weeks, while keeping
the three detectors at millikelvin temperatures. Before resuming the low-background run,
the ionization trigger was replaced by the phonon trigger, after a
thorough comparison of their relative performances. At the same time,
the online numerical filters on the center electrode signals were
adjusted to the new noise spectra, resulting in improved baseline
resolutions. \\
As a result, the varying quality of the data recorded
in the run 2003i required some selection, described in the following, 
while less than 0.5~\% of the data of the run 2003p (3.5 out of the 
1140~hours) had to be excluded because of data quality
cuts. In order to avoid biases, the data quality cuts are not made
event-by-event. Instead, the data quality is evaluated on an hourly
basis. \\
If one of the two following criteria fails, the entire hour
is rejected and deducted from the total exposure.
\begin{itemize}
\item{The first criterion is devised to
reject periods where the baseline noise of the heat channels reaches 
levels at which it reduces
significantly the nuclear recoil acceptance at low energy, for example
if the 3.29$\sigma$ $\gamma$-ray rejection removes events with $E_{R}$~$>$~30~keV.
In addition, this cut removes effectively periods where this noise
changes rapidly and the baseline resolution (and therefore the width
of the nuclear and electron recoil bands) cannot be evaluated reliably.
The FWHM baseline resolutions of the heat channel of the three detectors
are shown as a function of time in Fig.~\ref{fig:fig-datquali} and \ref{fig:fig-datqualp}
for the runs 2003i and 2003p, respectively. \\
The baseline resolution of a given
detector is calculated from the dispersion of amplitudes of events
where this detector did not participate in the trigger. It is evaluated for every
hour, with a three-hour averaging window. The dotted lines in Fig.~\ref{fig:fig-datquali}
represent times when the cryostat was re-filled with liquid He.
This procedure induces microphonic perturbations that persist for
hours, explaining most of the observed short episodes of degradation
of the baseline. These periods are removed by eliminating all hours
during which the average baseline deviates significantly from its typical
value. The cuts are set at 2.5, 5.0 and 1.0~keV$_{ee}$ for GSA3, GSA1
and GGA3, respectively. The detector GSA1 in the run 2003i was particularly
sensitive to microphonics. After 900~hours of low-background data taking,
the heat channel started to oscillate and it contaminated
its ionization channel and had to be removed from the trigger. Its
read-out electronics was switched off a few days later when it was
established that it also deteriorated the noise conditions of GSA3.
In the selected periods for GSA1, there are still important fluctuations
of the hourly average of the heat FWHM. In order to evaluate more
accurately the width of the nuclear recoil band for this sample, the
GSA1 data are divided into two subsets, according to whether the average
resolution is below or above 3~keV$_{ee}$ (named Quality 1 and Quality 2 
respectively). Consequently, two subsets
and two values of heat baseline FWHM appear in Table~\ref{tab:tab-Data-resos}
for GSA1 in the run 2003i. The FWHM cut effect on the data sets is
the following: out of the 1700~hours of the run 2003i, this cut removes
3.7~\%, 51.7~\% and 0.2~\% of the data set for GSA3, GSA1 and GGA3, respectively,
while no data are removed from the run 2003p.}
\item{The second criterion is that the drift in heat gain
be less than 10~\%, as discussed in the previous section. In the run
2003i, this cut removes 0.8~\% of the data for GGA3 and nothing for
the other two detectors. In the run 2003p, only one episode of 3.5~hours
is rejected out of 1140~hours of low-background run due to a failure
of the temperature regulation resulting in a drift exceeding 1~mK.}
\end{itemize}
To calculate the exposure in kg$\cdot$days, the
accepted hours are multiplied by the fiducial masses. The calculation
also takes into account the 6~\% loss due to regular shorting of the
electrodes and the dead-time losses. The fraction of dead time is
4~\% in run 2003p, and varies from 8~\% to 10~\% in the run 2003i, depending
on the detector.\\
In total, the fiducial volume data of the runs 2003i
and 2003p represent 25.7 and 22.7~kg$\cdot$d, respectively. The
data quality in the run 2003p is more uniform, as the nuclear recoil
bands of the three detectors have very similar widths in this configuration.
The three detectors remained extremely stable over the entire four-month
period that covered the run 2003p and the long calibration runs.
\subsection{\label{sec:simu}Simulation of WIMP detection efficiency}
\begin{figure}[tbp]
\includegraphics[scale=0.85]{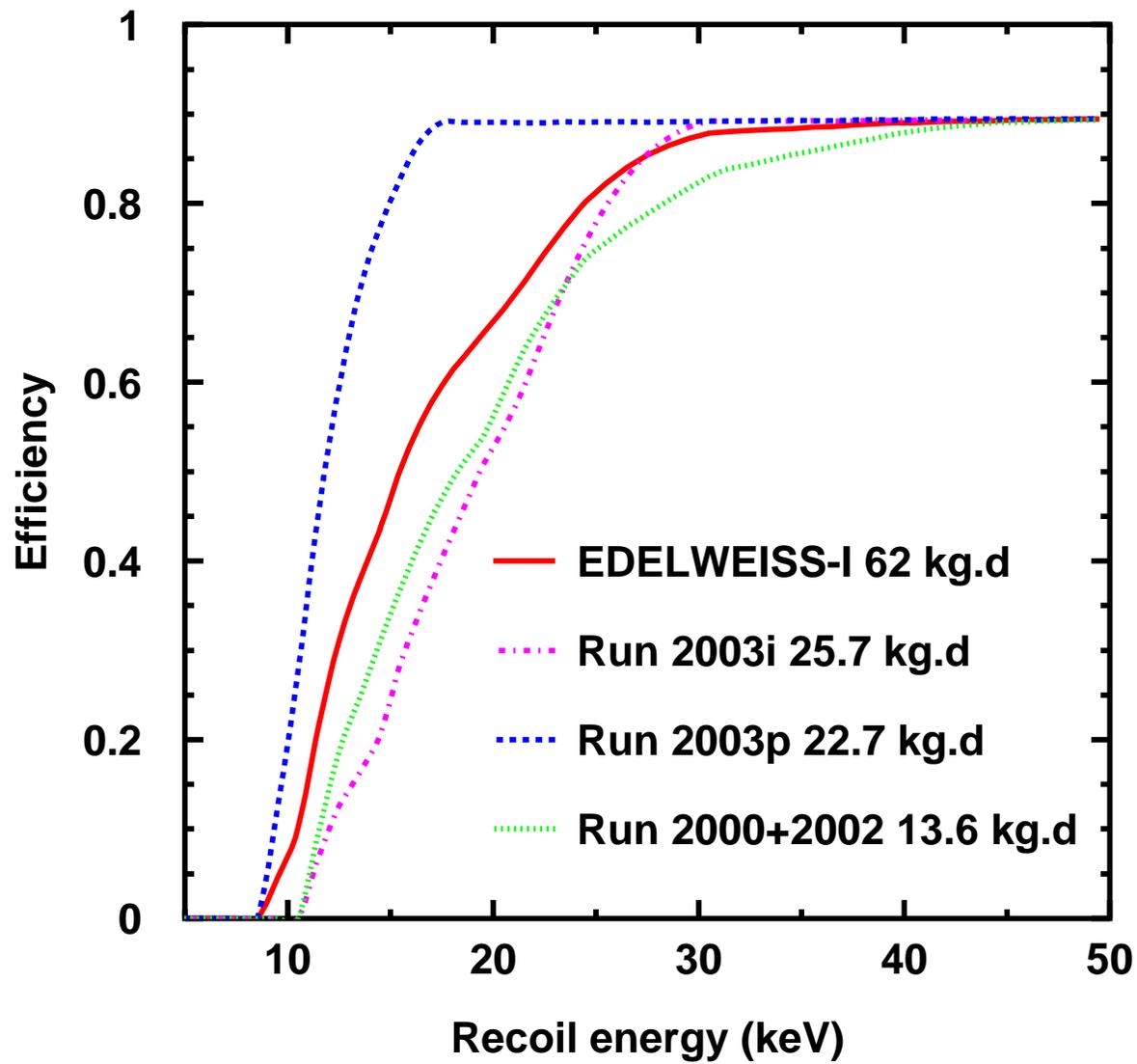}
\caption{\label{fig:fig-efficmc}Simulated efficiency, including all experimental
cuts and resolutions, as a function of recoil energy, calculated for
$M_{W}$~=~100~GeV/c$^{2}$, for runs 2000+2002, 2003i, 2003p and
the sum of all EDELWEISS-I runs.}
\end{figure}

In order to derive limits on the spin-independent
scattering rate of WIMPs in the detectors from the observed distribution
of events as a function of energy, it is necessary to take into account
the experimental efficiency. The experimental thresholds on ionization
energy and the resolution of the heat and ionization measurements
are inputs of a Monte Carlo simulation of the detector response to
WIMPs of given masses between 10~GeV/c$^{2}$ and 5~TeV/c$^{2}$.\\
The starting point of the simulation is with the analytical calculation
of the recoil energy spectrum using the formula and the prescriptions
of Ref.~\cite{bib-lewinsmith}. A spherical isothermal halo of WIMPs
with a local density of 0.3~GeV/c$^{2}$/cm$^{3}$ is assumed, with
a \emph{rms} velocity of 270~km/s,
an escape velocity of 650~km/s and an Earth-halo relative velocity
of 235~km/s. The spectrum is multiplied by the form factor for coherent 
scattering proposed in Refs.~\cite{bib-helm,bib-lewinsmith}.\\
This analytical recoil energy spectrum is then convolved
with the experimental resolutions. To do this, recoil events are simulated
with recoil energy values $E_{R}$ randomly distributed according
to the analytical calculation. The value of the quenching factor $Q$
is randomly chosen in a Gaussian distribution centered at $Q=0.16E_{R}^{0.18}$
with a \emph{rms} value $C$~=~0.035 
(see Sec.~\ref{sec:calibration}). $E_{R}$ and $Q$ are converted into ionization and heat
signals using the inverse of Eqs.~\ref{eq:Er} and \ref{eq:Q}. The
ionization and heat signals are then independently smeared using the measured 
resolutions at 0 and 122~keV listed in Table~\ref{tab:tab-Data-resos}, 
interpolated using the method of Ref.~\cite{bib-edw-calib}.\\
The smeared values for the recoil energy $E_{R}^{*}$
and quenching factor $Q^{*}$ are calculated from the smeared ionization
and heat signals $E_{I}^{*}$ and $E_{H}^{*}$ using Eqs.~\ref{eq:Er}
and \ref{eq:Q}. The simulated data are then subjected to the same
cuts as the physics data, namely: the cut on the ionization energy,
the 1.65$\sigma$ selection of nuclear recoils and the 3.29$\sigma$
rejection of electron recoils.\\
These calculations are repeated for the eleven configurations
listed in Table~\ref{tab:tab-Data-sets}, normalized to the corresponding
exposure and summed. The result is a predicted energy spectrum for each 
WIMP mass for the entire EDELWEISS-I data set. The limits on the 
WIMP-nucleon scattering cross-section
as a function of WIMP mass are obtained by comparing directly these
predicted spectra to the data (see Sec.~\ref{sec:limite}).\\
To check the validity of the simulation, an efficiency
as a function of recoil energy is calculated by dividing the predicted
spectrum by the result of the analytical calculation. For a given
run configuration, this curve can be compared to the results of the
efficiency measurement performed with neutron coincidences. 
In Table~\ref{tab:tab-Data-sets}
are compared the simulated and measured energies at which half of
the maximal trigger efficiency is reached. The simulation agrees with the
measured values to within 1~keV. No measurement is available for
GeAl6 because of the absence of neutron coincidence data. For GSA1, 
there is only one measurement because the neutron calibration was done 
when the heat baseline resolution was 2.4~keV$_{ee}$, corresponding 
to the "Quality 1" configuration (see Sec.~\ref{sec:data-new}).\\
The simulated efficiencies 
as a function of recoil energy for the entire EDELWEISS-I data set,
and also separately for the runs 2000+2002, 2003i and 2003p are shown
in Fig.~\ref{fig:fig-efficmc}, for a WIMP mass of 100~GeV/c$^{2}$. \\
The significant increase in efficiency
at low energy obtained with the phonon trigger configuration is clearly
displayed. For the entire data set, the efficiency reaches half of
its maximal value at 15~keV, and 75~\% at 20~keV.
\section{\label{sec:results}Results and discussion}
\subsection{\label{sec:exp-data}Experimental results}
\begin{figure*}[tbp]
\mbox{
\includegraphics[scale=0.4]{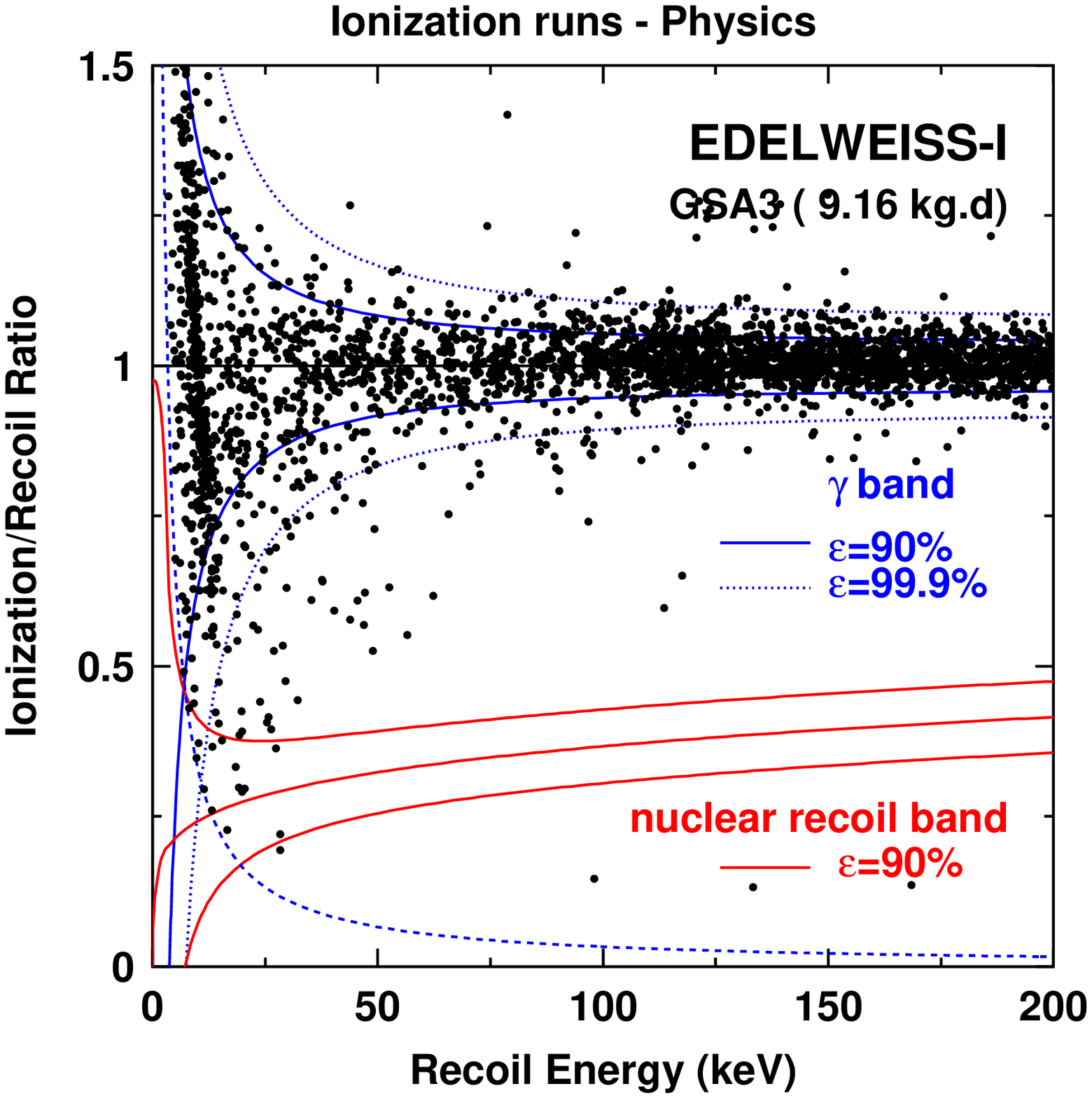}
\includegraphics[scale=0.4]{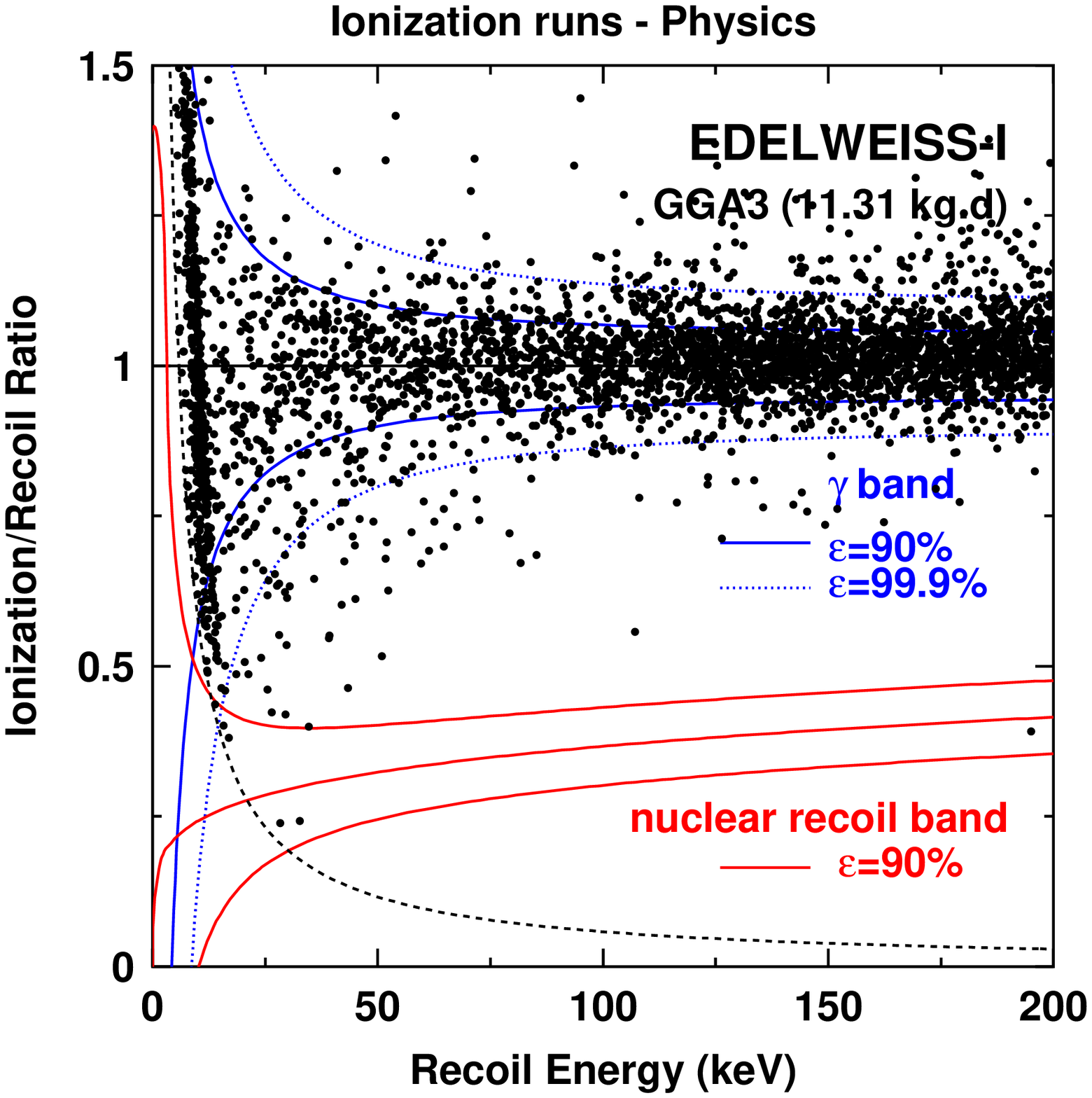}}
\caption{\label{fig:fig-ion1-3}Distribution of the quenching
factor $Q$ (ratio of the ionization to the recoil energies) as a
function of the recoil energy $E_{R}$ for data collected in the fiducial
volume of GSA3 and GGA3 in the run 2003i. Also plotted as full lines
are the $\pm$~1.65$\sigma$ (90~\%) electron and nuclear recoil bands.
The dotted lines represent the $\pm$~3.29$\sigma$ (99.9~\%) electron
recoil band. The hyperbolic dashed curve corresponds to the ionization
energy threshold.}
\end{figure*}

\begin{figure*}[tbp]
\mbox{
\includegraphics[scale=0.4]{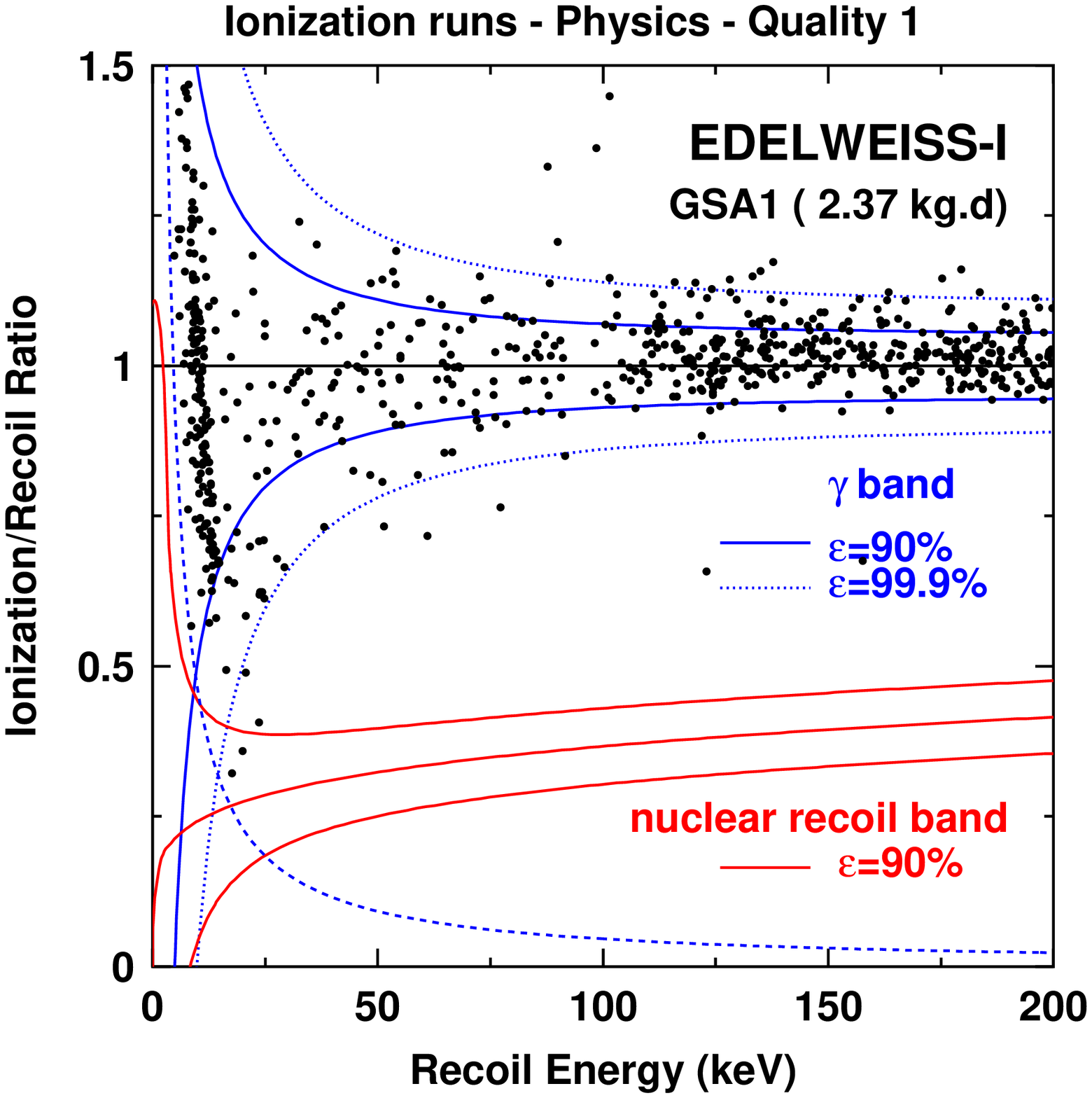}
\includegraphics[scale=0.4]{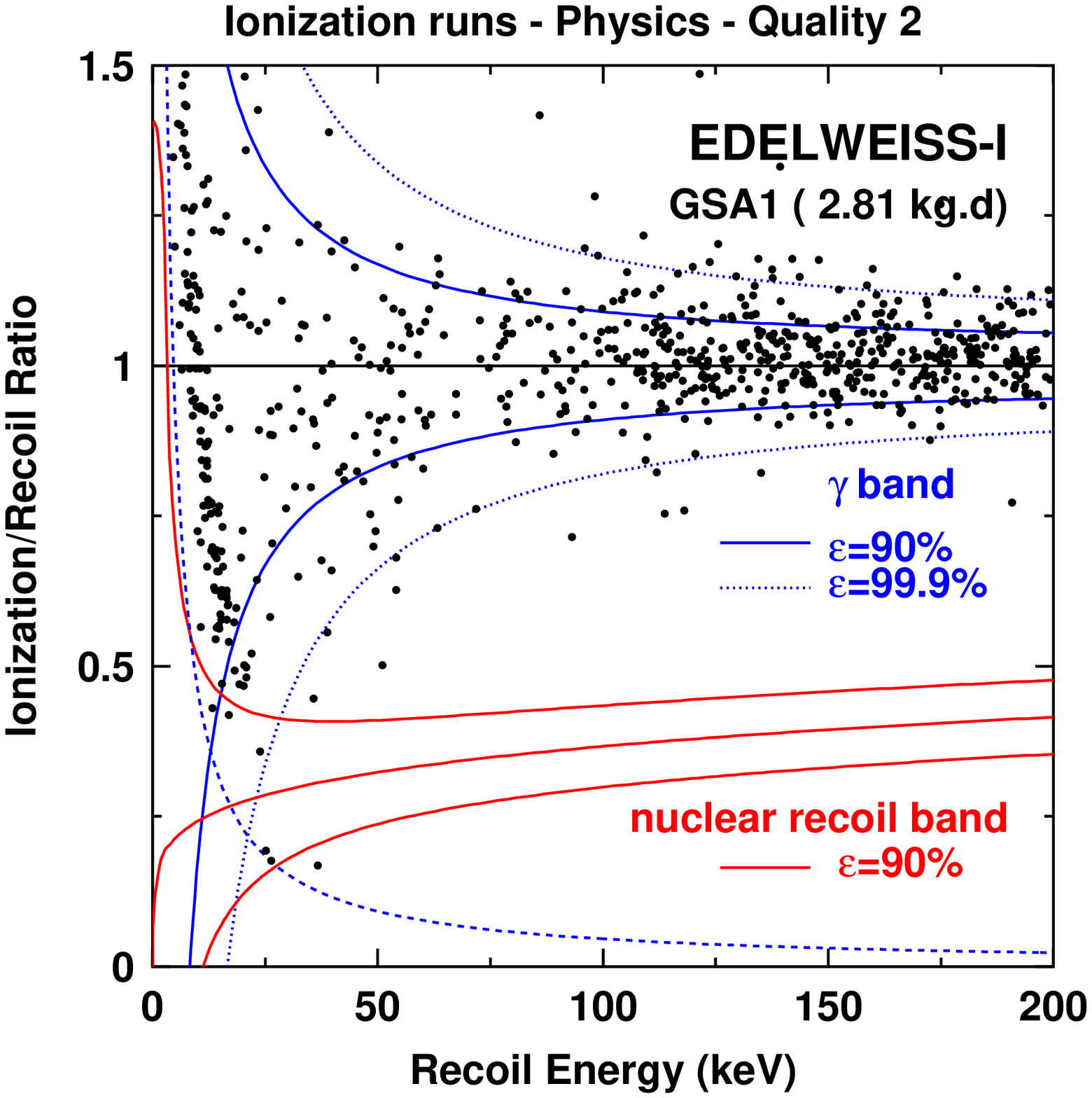}}
\caption{\label{fig:fig-ion2}Same as previous figure, for GSA1
in the run 2003i. Because of important fluctuations of the hourly
average of the heat FWHM resolution, the data recorded with this detector
are divided into two subsets according to whether this value is below
3~keV$_{ee}$ (Quality 1) or between 3 and 5~keV$_{ee}$ (Quality
2).}
\end{figure*}

\begin{figure*}[tbp]
\mbox{
\includegraphics[scale=0.4]{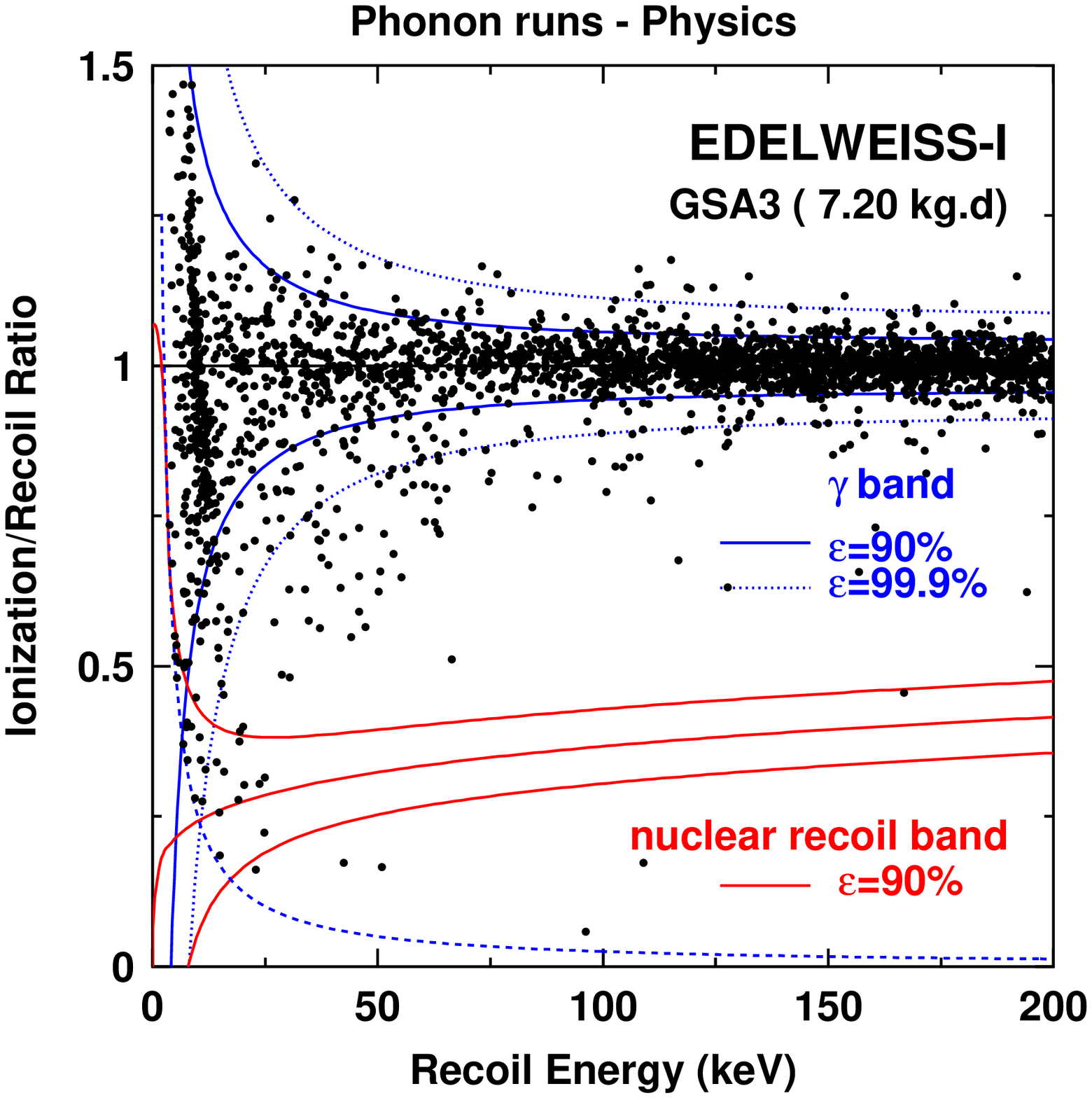}
\includegraphics[scale=0.4]{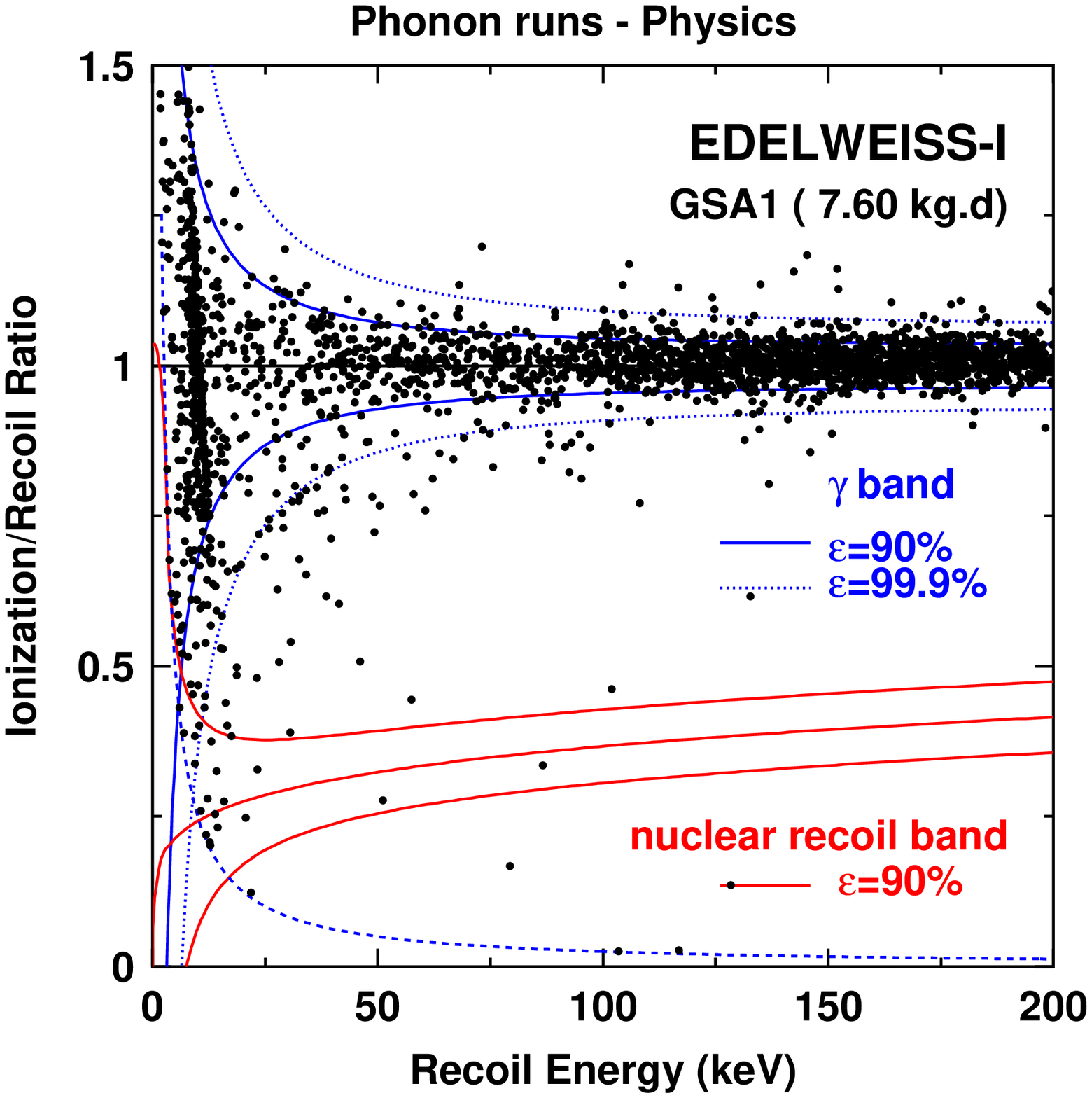}}
\caption{\label{fig:fig-phon1-2}Same as previous figure, for
GSA3 and GSA1 in the run 2003p.}
\end{figure*}

\begin{figure*}[tbp]
\mbox{
\includegraphics[scale=0.4]{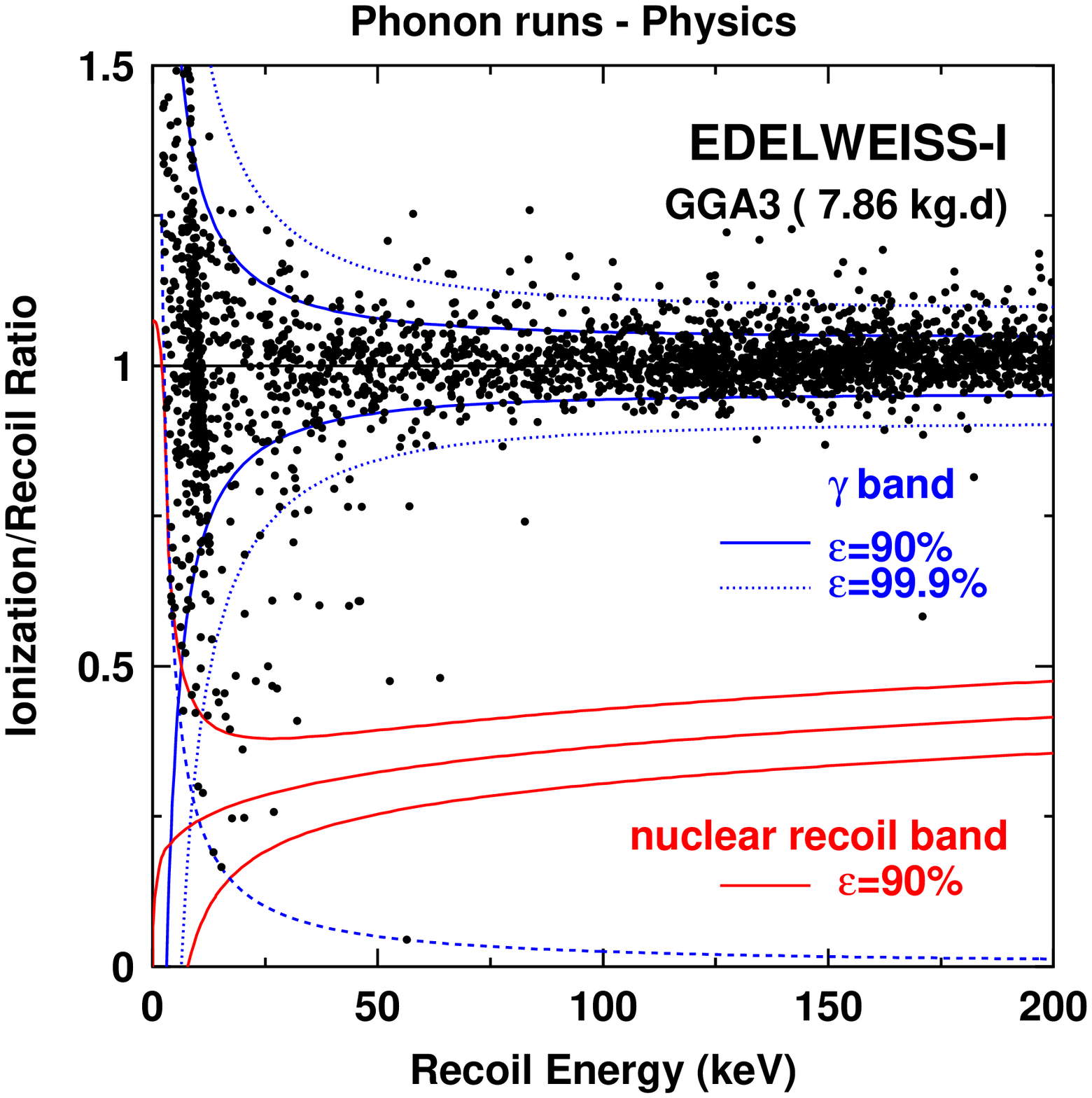}
\includegraphics[scale=0.4]{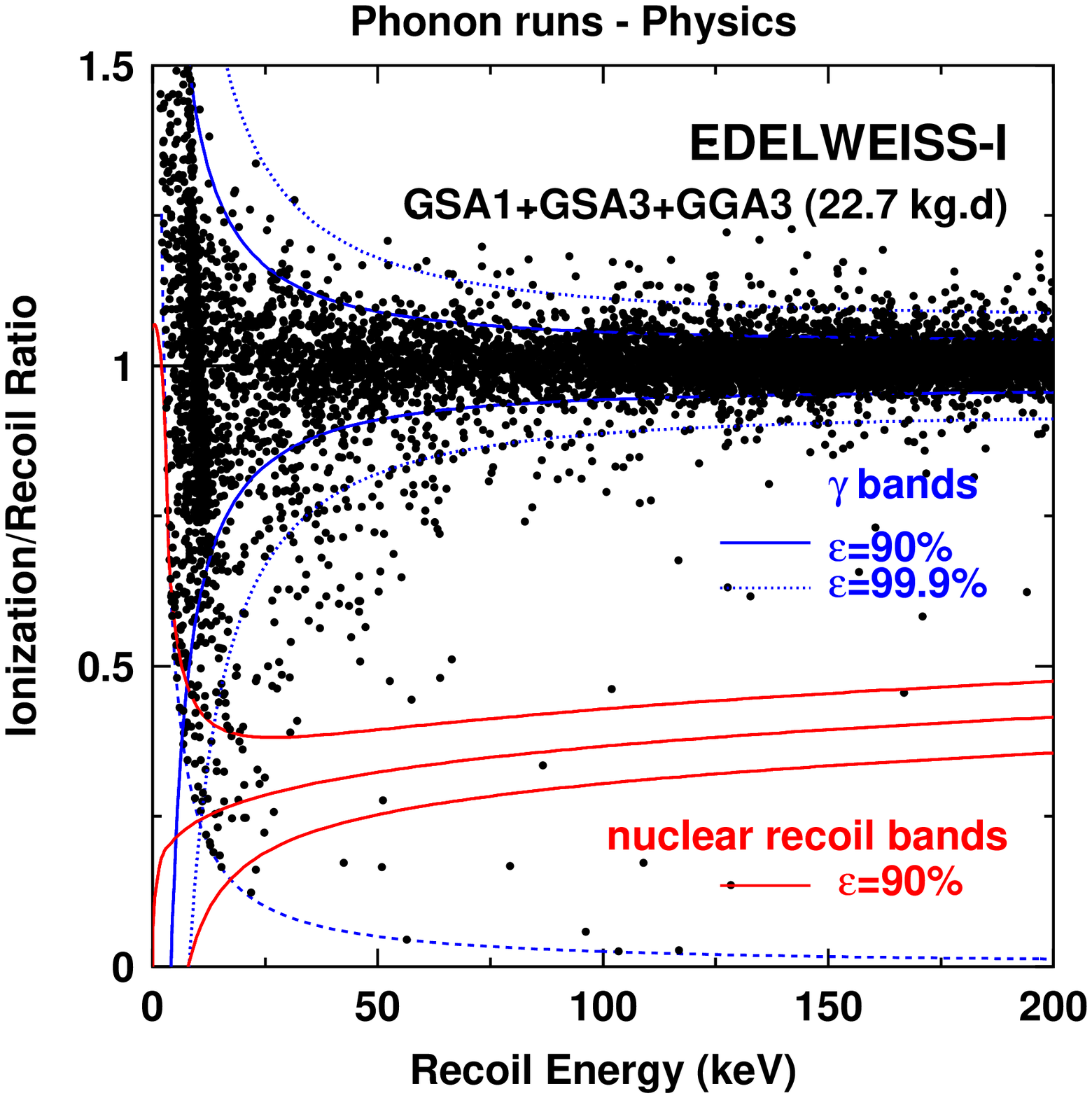}}
\caption{\label{fig:fig-phon3-all}Same as previous figure,
for GGA3 and for the sum of the three detectors in the run 2003p.}
\end{figure*}

For the run 2003p, the event rate in the total volume of the three
detectors before the nuclear recoil selection and $\gamma$-ray
rejection is 2.00~$\pm$~0.03~evt/keV/kg/d between 30 and 100~keV.
The fiducial volume selection reduces the raw rate to 1.31~$\pm$~0.03~evt/keV/kg/d
in the same energy range. A significant fraction of these events are
coincidences between detectors: the single rate is 0.98~$\pm$~0.03~evt/keV/kg/d.
Most of these events are electron recoils as can be seen in 
Figs.~\ref{fig:fig-ion1-3} to~\ref{fig:fig-phon3-all}
showing the distributions of $Q$ as a function of $E_{R}$ for the
runs 2003i and 2003p. The corresponding figures for the runs 2000
and 2002 can be found in Refs.~\cite{bib-edw2000} and~\cite{bib-edw2002},
respectively. \\
In total, EDELWEISS-I has accumulated 62~kg$\cdot$d
of fiducial volume data. The recoil energy spectrum of all the events
passing the nuclear recoil selection described in Sec.~\ref{sec:selection}
is shown in Fig.~\ref{fig:fig-Recoil}.\\ 
Most counts are below 30~keV
(53 counts), only three are between 30 and 100~keV, and three more are
between 100 and 200~keV. The average count rate between 30 and 200~keV
is 6$\times10^{-4}$~counts/keV/kg/d. Sixteen counts are observed
between 20 and 30~keV, and eighteen more between 15 and 20~keV.
The few counts observed below 15~keV must be interpreted with care,
as the efficiency drops rapidly in this region. This drop explains
the low-energy shape of the simulated WIMP spectra shown on the same
figure, calculated for different WIMP masses and an arbitrary choice
of the spin-independent WIMP-nucleon scattering cross-section 
$\sigma_{W-n}$~=~10$^{-5}$~pb. Indeed, as it will be shown later, 
the range below 15~keV is not used for setting limits on $\sigma_{W-n}$.
\subsection{\label{sec:compatibility}Compatibility between the different data sets.}
To check whether the 2000+2002, 2003i and 2003p
data sets are compatible and can be added, the following test has
been performed. First, the total spectrum of Fig.~\ref{fig:fig-Recoil}
has been corrected for the total efficiency for the 62~kg$\cdot$d,
as calculated by the simulation and shown in Fig.~\ref{fig:fig-efficmc}.\\
Then, in Fig.~\ref{fig:fig-compatible}, this corrected spectrum is alternatively
multiplied by the simulated efficiencies of the runs 2000+2002, 2003i
and 2003p and compared with the corresponding data set. As no significant
deviations from the average behavior are observed above 15~keV, it is justified to
add the three data sets together. The factor four increase in exposure
and the significant increase in efficiency at low energy explains
why 16 events are observed between 20 and 30~keV in the new data
set while none were reported in Refs.~\cite{bib-edw2000,bib-edw2002}. Conversely,
the few events observed just below the 20 and 30~keV analysis thresholds
in Refs.~\cite{bib-edw2000,bib-edw2002} are consistent with the expectations
deduced from the new data set recorded with a much better efficiency
at low energy.
\subsection{\label{sec:int-exp-data}Data interpretation}
\begin{figure}[tbp]
\includegraphics[scale=0.85]{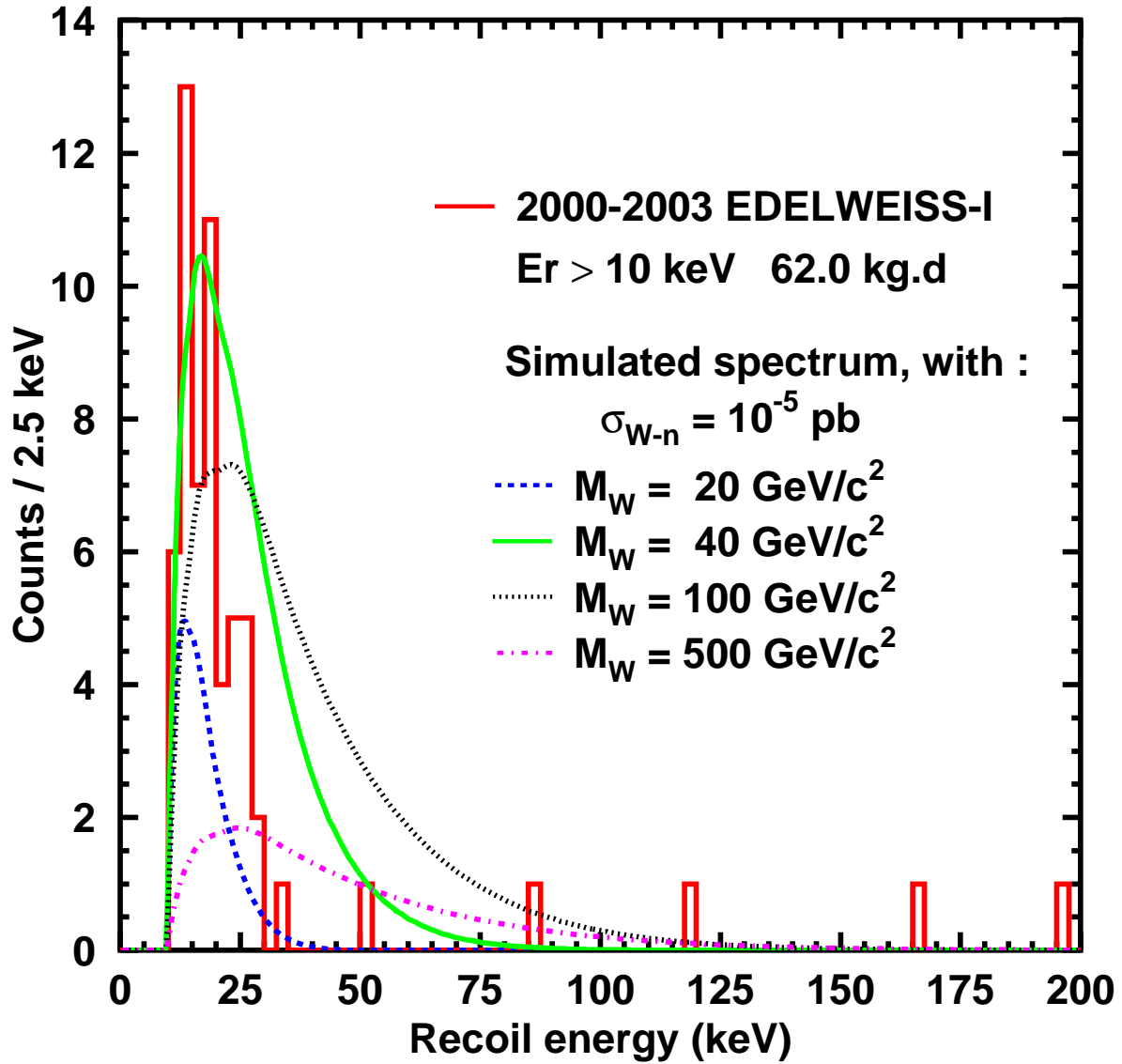}
\caption{\label{fig:fig-Recoil}Recoil energy spectrum of events in the nuclear
recoil selection ($E_{R}$~$>$~10~keV), recorded by EDELWEISS-I for
a total fiducial exposure of 62~kg$\cdot$d, compared with simulated
WIMP spectra using a WIMP-nucleon scattering cross-section 
$\sigma_{W-n}$~=~10$^{-5}$~pb
for WIMP masses $M_{W}$~=~20,~40,~100 and 500~GeV/c$^{2}$.}
\end{figure}

\begin{figure}[tbp]
\includegraphics[scale=0.85]{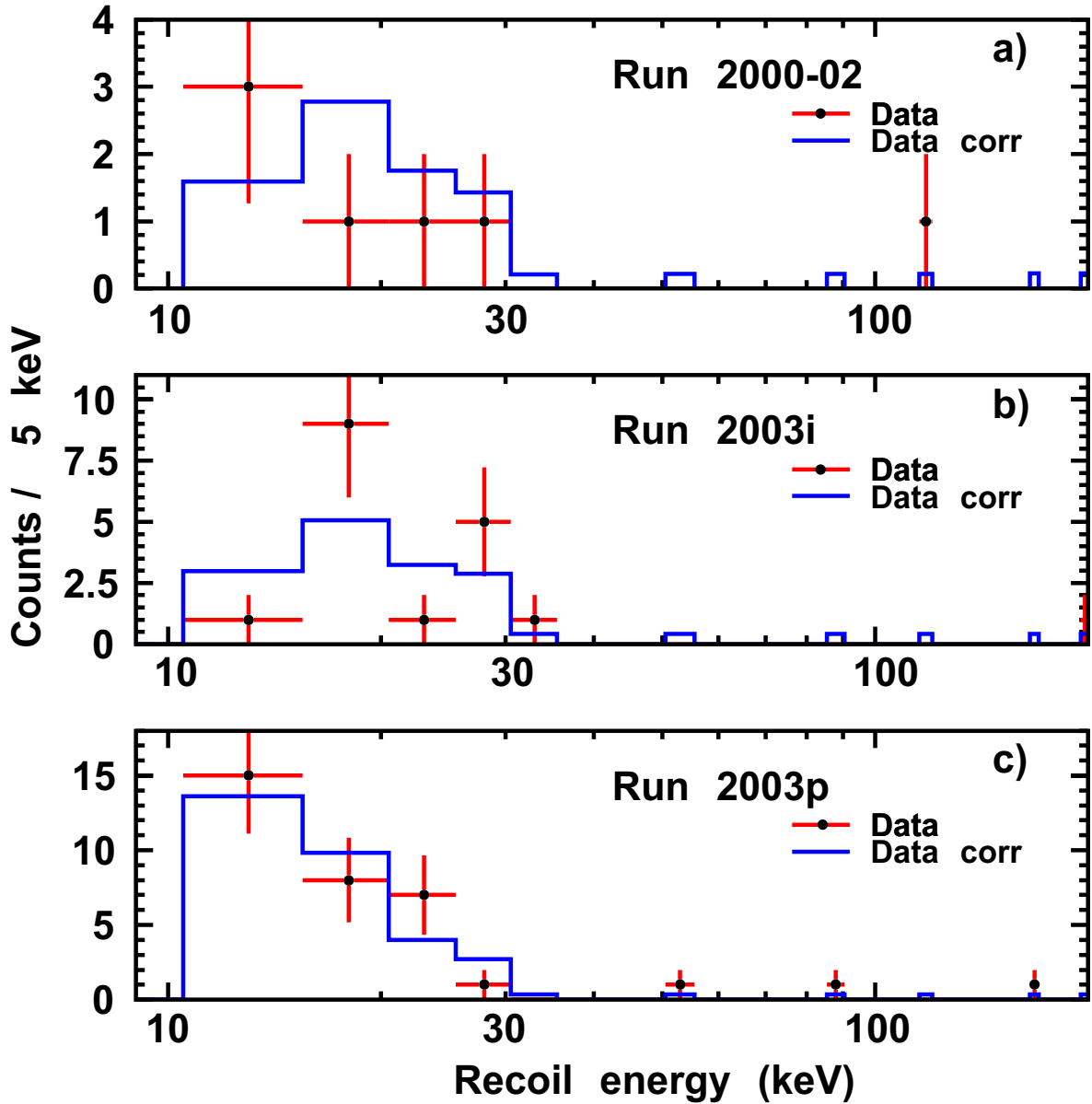}
\caption{\label{fig:fig-compatible}Data points: recoil energy spectrum of events
in the nuclear recoil selection recorded by EDELWEISS-I in the (a)
2000 and 2002, (b) 2003i and (c) 2003p runs. These spectra are compared
with the efficiency-corrected average spectrum recorded in the entire
data set (full-line histogram), obtained by multiplying the experimental
spectrum of Fig.~\ref{fig:fig-Recoil} by the ratio of energy-dependent
efficiencies for the run of interest and for the entire data set. }
\end{figure}

\begin{figure}[tbp]
\includegraphics[scale=0.85]{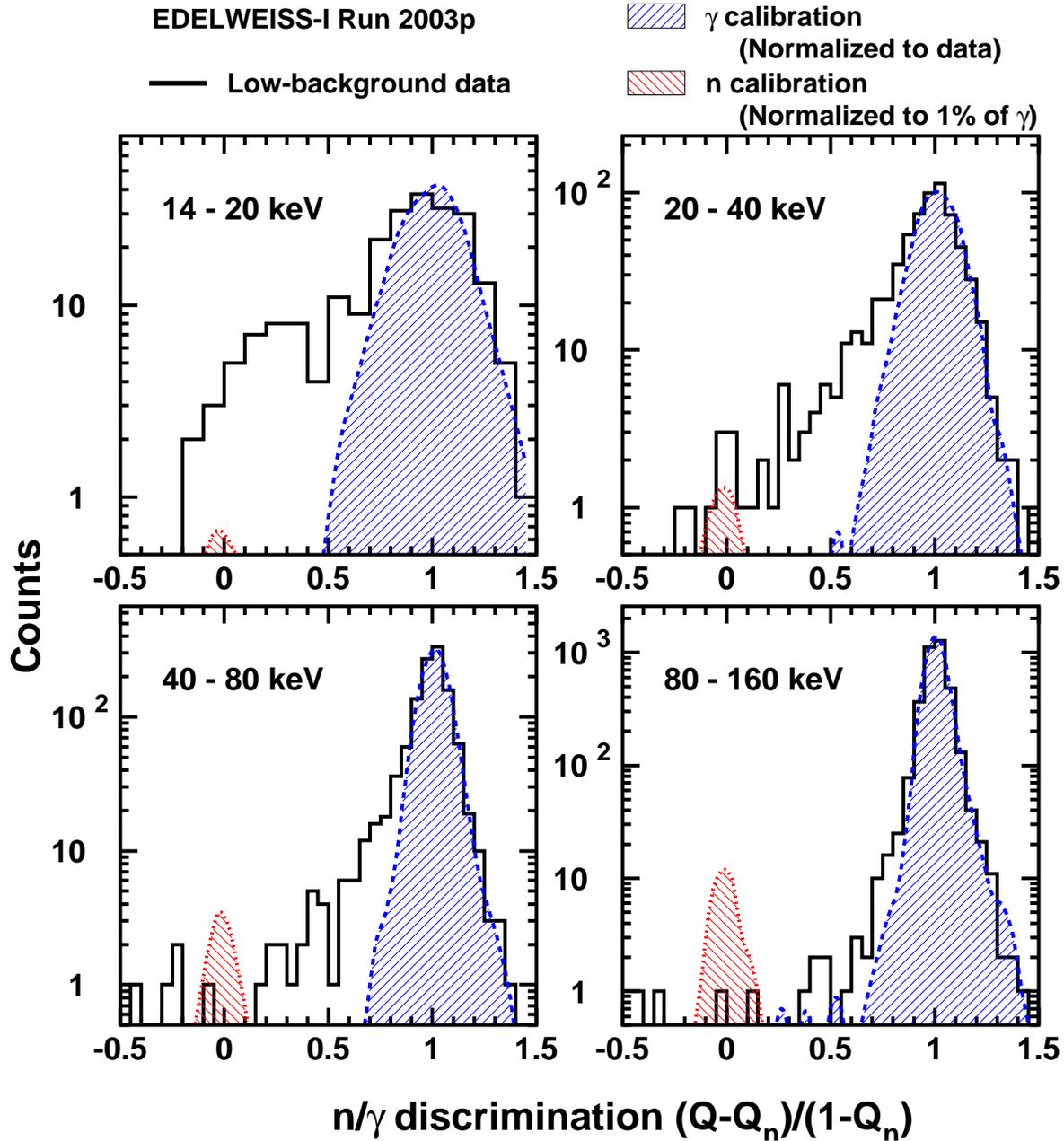}
\caption{\label{fig:fig-qsli}Distribution of the normalized
quenching $D=(Q-Q_{n})/(1-Q_{n})$ for four intervals of recoil energy
(14-20, 20-40, 40-80 and 80-160~keV) for the three detectors in the
run 2003p. With this variable, nuclear and electron recoils are centered
at 0 and 1, respectively. Full line histogram: low-background run. Hatched 
distribution centered at 1: high-statistics $\gamma$-ray calibration ($^{137}$Cs
source) normalized to the area of the upper half (Q~$>$~1) of the
$\gamma$-ray peak in the low-background run. Hatched
distribution centered at 0: high-statistics neutron source calibration ($^{252}$Cf)
normalized to 1~\% of the area of the $\gamma$-ray peak.}
\end{figure}

\begin{figure}[tbp]
\includegraphics[scale=0.85]{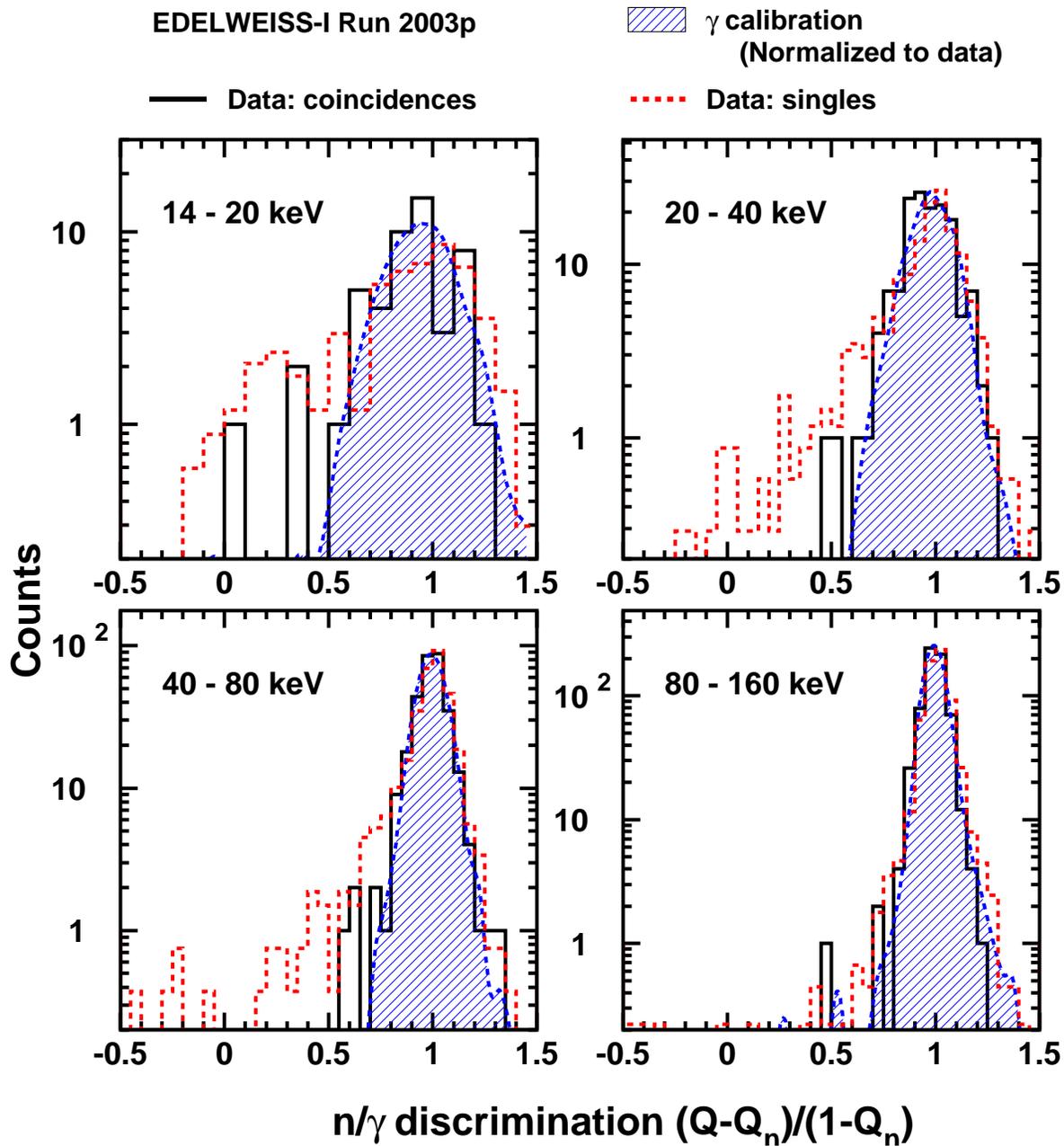}
\caption{\label{fig:fig-Qcoinc}Same as previous figure, except for coincidence
(full line histogram) and single-detector (dashed line histogram)
events. Hatched distribution: high-statistics $\gamma$-ray
calibration ($^{137}$Cs source). All spectra are normalized to the
area of the upper half (Q~$>$~1) of the $\gamma$-ray peak of the coincidences
in the low-background run. }
\end{figure}

In the following section, the experimental spectrum
of Fig.~\ref{fig:fig-Recoil} will be interpreted in terms of a 90~\%
C.L. limit on the WIMP scattering rate without subtracting any background.
However, it is clear from the comparison with the simulated WIMP spectra
that no single WIMP mass can explain the entire spectrum. This suggests
that part of the spectrum may be attributed to non-WIMP background.
As it will be shown in this section, this conclusion can also be reached
independently by studying the behavior of the data lying just above
the nuclear recoil band, and by studying coincidences between the
detectors.\\
Fig.~\ref{fig:fig-qsli} shows the distributions of the normalized quenching 
\begin{equation}
D=\frac{Q-Q_{n}(E_{R})}{1-Q_{n}(E_{R})}
\label{eq:deq}
\end{equation}
 for the data recorded by the three detectors in the run 2003p, for
three intervals of recoil energy.\\
 With this variable, where 
$Q_{n}(E_{R})=0.16E_{R}^{0.18}$,
nuclear and electron recoils should appear as peaks centered at 0
and 1, respectively, independent of $E_{R}$. Indeed, 
the superposed hatched histograms represent the distributions recorded
in neutron and $\gamma$-ray calibrations. The $\gamma$-ray calibration
 data are normalized to have the same number of counts above $D$~=~1
as in the low-background run. This clearly shows that the latter distribution
is not symmetric around $D$~=~1 as it is in $\gamma$-ray calibrations.
In the low-background run, the tail extends down to $D$~=~0, especially
at low recoil energy. Below $E_{R}$~=~40~keV, the tail reaches down
to the region where neutrons and WIMPs are expected. This is close
to the energy below which the event rate in the nuclear recoil band
increases rapidly (see Fig.~\ref{fig:fig-Recoil}). This type of tail
in $D$ (and thus in $Q$) is generally attributed to bad charge collection of electron
recoils near the surface of the detector~\cite{bib-shutt}. \\
As seen in
Fig.~\ref{fig:fig-Qcoinc}, this tail is significantly reduced when requiring
a coincidence between detectors. This suggests that the mean free
path of the radiation at the origin of the events in the tail is less
than the 2~mm of Cu that separates two neighboring detectors. However,
the precise shape of the tail is not known and is difficult to predict,
especially near the nuclear recoil band. Therefore, no attempts have been
made to subtract a background contribution from this source to the observed rate in
the nuclear recoil band.\\
The study of coincidences between the detectors
in the low-background run gives a robust evidence for another possible
source of background: a residual neutron flux. One coincident event
between two nuclear recoils is observed between the fiducial volume
of GGA3 ($E_{R}$~=~15~keV, $Q$~=~0.27) and the outer volume in the
neighboring detector GSA1 ($E_{R}$~=~14~keV, $Q$~=~0.28). In Ref.~\cite{bib-cdms-suf}, 
is presented the case where two apparent nuclear recoils in neighboring
detectors are due to the coincidence of two surface electrons with
both charge collections being at the lower end of the tail in $Q$.
This process is unlikely in the EDELWEISS-I geometry, where the 2~mm
of Cu separating the detectors should prevent a single electron
or an electron cascade to touch two detectors. 
Indeed, Fig.~\ref{fig:fig-Qcoinc}
clearly demonstrates that the coincident events are mainly associated
with good charge collection down to low recoil energy, and, conversely,
bad charge collections are associated with single events. The observed
coincidence is very likely due to two coincident neutron interactions.
Monte Carlo simulations of the neutron background based on the measured
neutron flux at the LSM tend to predict single event rates of the
order of 1 event per 62~kg$\cdot$day, but the uncertainty on the
absolute scale is large. The simulations also predict that the ratio
of singles to coincidences for neutron scatters is approximately 10
to 1, for neutrons from the rock radioactivity. This is also the ratio
measured in neutron calibrations with a $^{252}$Cf source. It is
thus possible that some of the events in Fig.~\ref{fig:fig-Recoil} are
due to a residual neutron background. On the other hand, given a rate
of one neutron per 62~kg$\cdot$day and a probability for this event
to be a coincidence of the order of 10~\%, it is also possible that none
of the single events are neutrons.\\
A close inspection of the right panel of Fig.~\ref{fig:fig-phon3-all}
suggests a third possible source of background. There is an accumulation
of events along the hyperbola corresponding to the ionization threshold of 
2.5~keV$_{ee}$.
This may indicate that the cuts (see Sec.~\ref{sec:processing}) do not remove all NTD-events. However
the recoil energy of most of these events appearing in the nuclear
recoil band is below 15~keV and they do not affect significantly
the derived WIMP exclusion limits for WIMP masses above 25~GeV/c$^2$ 
(see Sec.~\ref{sec:limite}).\\
In summary, studies based on independent data sets
confirm that two sources of background may contribute significantly
to the observed rate in the nuclear recoil band: surface electrons
and neutrons. In the absence of more detailed studies, it is not possible
to conclude quantitatively and therefore no background subtraction
is performed for the estimate of the limits on the WIMP collision
rate in the detectors.
\subsection{\label{sec:limite}Neutralino scattering cross-section limits}
\begin{figure}[tbp]
\includegraphics[scale=0.85]{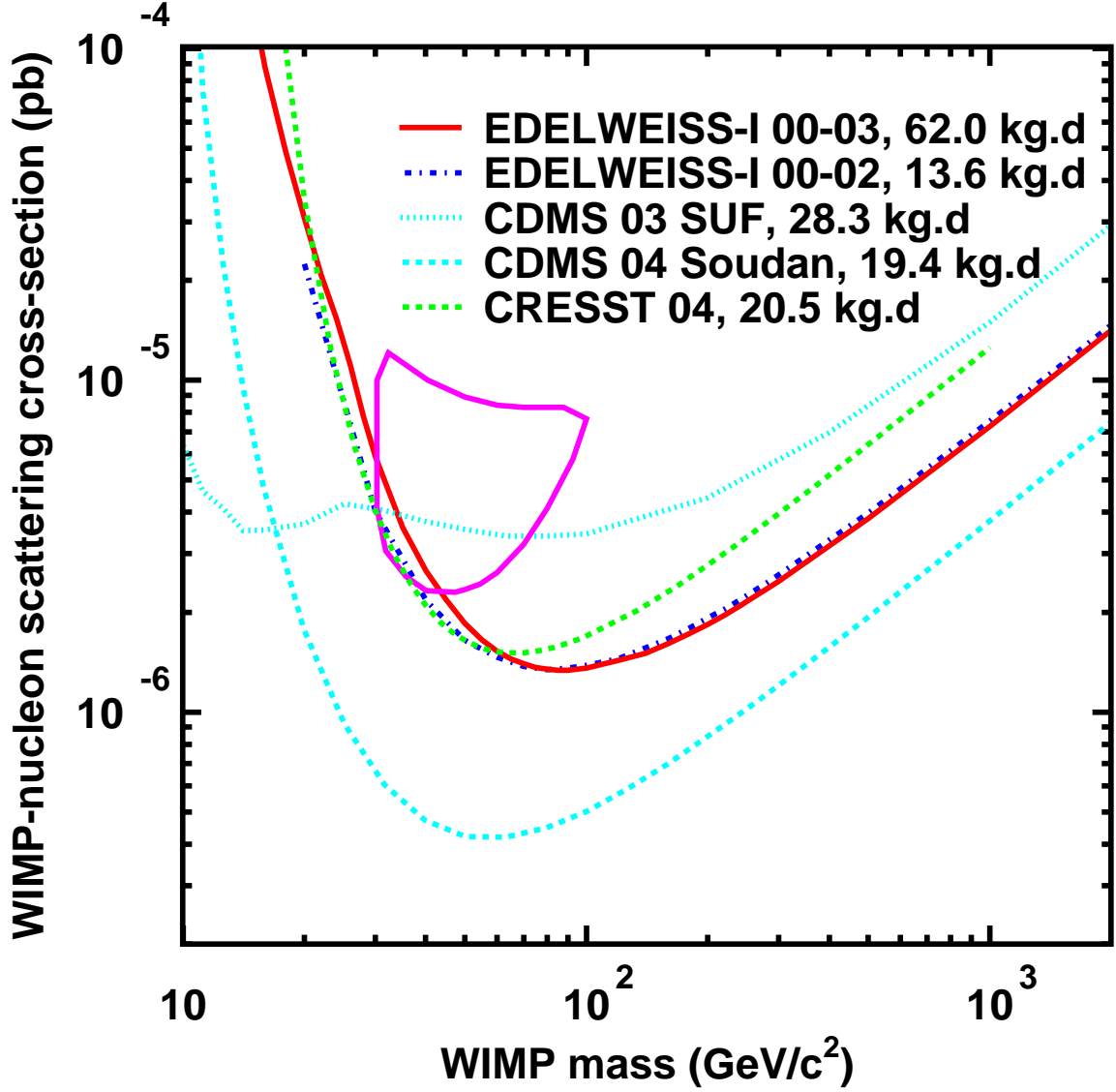}
\caption{\label{fig:fig-EDW-Limits}90~\% C.L. spin independent
limits (solid curve) obtained by EDELWEISS-I for a total fiducial
exposure of 62~kg$\cdot$d, for $E_{R}$~$>$~15~keV. Dotted curve
: 2003 CDMS limits~\cite{bib-cdms-suf}. Light dashed curve : 2004 CDMS
limits~\cite{bib-cdms-soudan}. Dark dashed curve : CRESST limits using 
W recoils~\cite{bib-cresst}.
Dash-dotted curve : previous published EDELWEISS-I limits~\cite{bib-edw2002}.
Closed contour : allowed region at 3$\sigma$ C.L. from the DAMA 1-4
annual modulation data~\cite{bib-dama}.}
\end{figure}

\begin{figure}[tbp]
\includegraphics[scale=0.85]{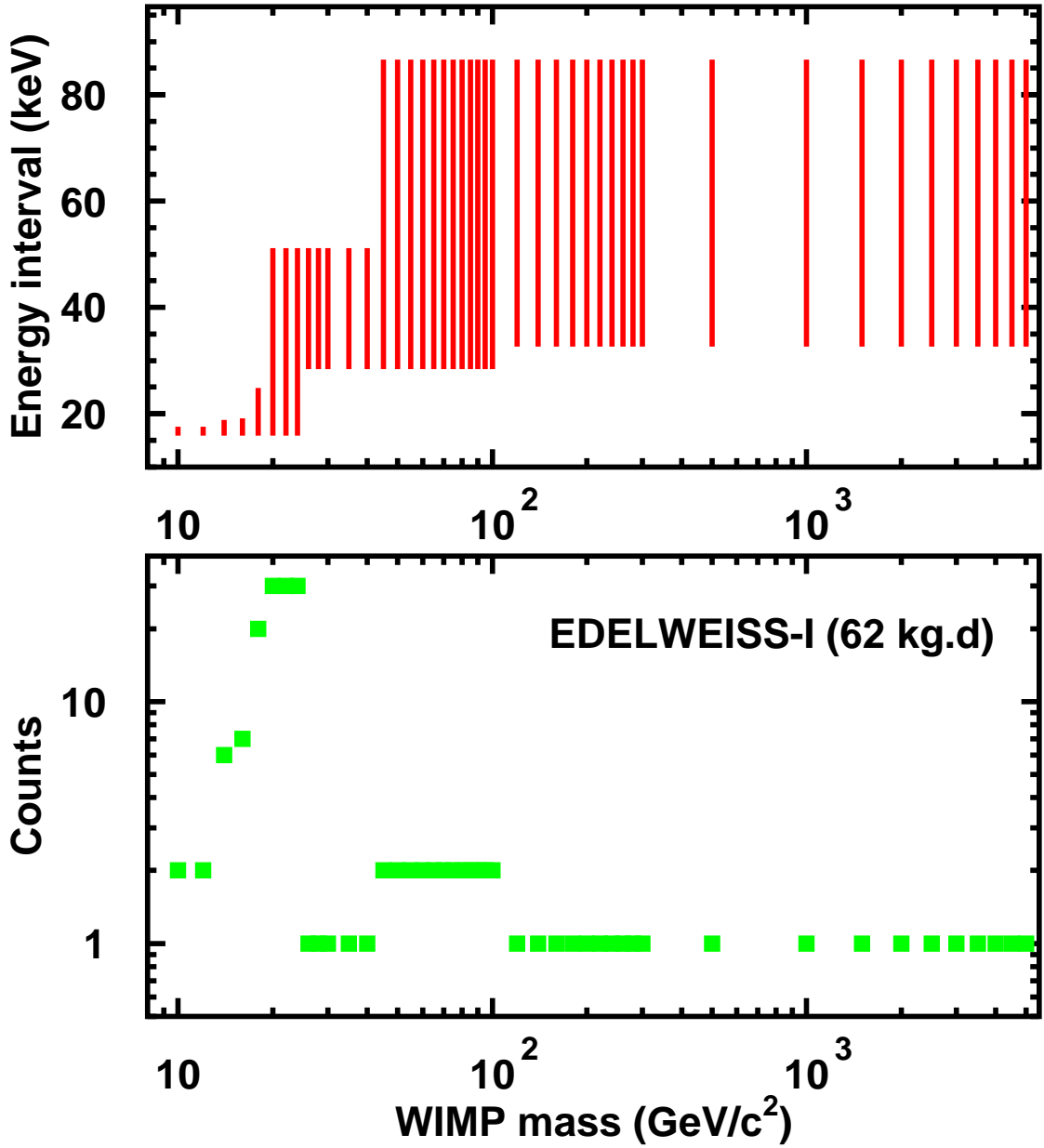}
\caption{\label{fig:fig-EDW-range}Top: Recoil energy range selected by the Yellin
algorithm used to derive the EDELWEISS-I 90~\% C.L. limit from its
62~kg$\cdot$d fiducial data set, as a function of WIMP mass. Bottom:
Number of events observed in the corresponding recoil energy range.}
\end{figure}

In order to set upper limits on the cross-section
of the spin-independent scattering of a WIMP on a nucleon $\sigma_{W-n}$
as a function of the WIMP mass $M_{W}$, the optimum interval method of 
Ref.~\cite{bib-yellin} is used. This method is well adapted to the present case,
where no reliable models are available to describe potential background sources 
and no subtraction is possible. This method can be summarized in the
following way: for each mass $M_{W}$, the upper limit on $\sigma_{W-n}(M_{W})$
is calculated using the number of events observed in the recoil energy
interval that provides the strongest constraint. Of course, the 90~\%
C.L. limits on $\sigma_{W-n}(M_{W})$ that would be derived from these
carefully chosen intervals by using simple Poisson statistics would
be biased and too optimistic. Using Monte Carlo simulations, these biases 
have been precisely tabulated in~\cite{bib-yellin} in such a way that they can be
corrected for and thus derive reliable 90~\% C.L. limits. This method
automatically determines which energy interval provides the strongest
constraint on the presence of a signal. This energy interval may contain
events. No background is subtracted, and indeed in the presence of
a background having the same energy spectrum as the WIMP signal, the
derived 90~\% C.L. limit is similar to the Poisson limit based on the
total number of observed events in the entire interval.\\
The inputs of the method are \emph{i)} the individual recoil energies of the nuclear recoil 
candidates (see Fig.~\ref{fig:fig-Recoil}) and \emph{ii)} the
expected recoil energy spectra for WIMPs, calculated using the simulation
described in Sec.~\ref{sec:simu}, as a function of WIMP mass. We
use $E_{R}$~$>$~15~keV, corresponding to the recoil energy where
the efficiency reaches half of its maximal value. The resulting 90~\%
C.L. limits on $\sigma_{W-n}(M_{W})$ are shown in Fig.~\ref{fig:fig-EDW-Limits}.\\
As this method determines the energy interval
that constrains the most the signal, this information provides some
assistance in the interpretation of the observed spectrum. \\
The lower
and upper bounds of the selected energy intervals are shown in 
Fig.~\ref{fig:fig-EDW-range},
together with the number of events in the corresponding intervals.
For $M_{W}$~$>$~25~GeV/c$^{2}$, the selected intervals are in the
energy range from 28.4 to 86.6~keV. This corresponds well to what
is observed in Fig.~\ref{fig:fig-Recoil}, where the experimental spectrum
is compared with the expected signal for WIMPs of different masses
and an arbitrary choice of $\sigma_{W-n}=10^{-5}$~pb. Most of the
observed events are below $E_{R}$~=~30~keV. In contrast, 
for $M_{W}$~$>$~20~GeV/c$^{2}$,
a significant part of the recoil spectra lies above this energy. For
masses in the 20~-~25~GeV/c$^{2}$ range, the spectrum is strongly
peaked below 30~keV, and the experimental data provide a much weaker
constraint on $\sigma_{W-n}$. For this WIMP mass interval, 
the best limits are obtained from the 30 events in the energy range
from 15.9 to 51.1~keV and are similar to the corresponding Poisson
limits.\\
When comparing the limits shown in Fig.~\ref{fig:fig-EDW-Limits} with theoretical predictions, 
one should take into account the large theoretical uncertainties associated
with the astrophysical and nuclear model parameters.
These were chosen according to the presciptions of Ref.~\cite{bib-lewinsmith}
that provide a framework for comparing the sensitivities
of the different experiments.
The experimental systematic uncertainties on the present limits that are
relevant for this kind of comparison have been studied.
Since the results rely heavily on the recoil energy interval between
28.4 and 86.6~keV, the uncertainties
on the energy threshold ($\pm$~1~keV, see Table~\ref{tab:tab-Data-sets})
and on the NTD event cut ($<$~0.1~\% down to 10~keV, Sec.~\ref{sec:processing})
have a negligible influence.
More important are the contribution from the determination
of the fiducial volume and 
of the position and width of the nuclear recoil band.
These effects are discussed in Ref.~\cite{bib-edw-calib}.
Here, they both correspond to $\sim$~2~\% uncertainties on the
experimental efficiency in the relevant energy range,
although these may be overestimations since conservative
choices are made in the experimental determination of these
quantities~\cite{bib-edw-calib}.
In addition, the 1~\% uncertainties on the absolute ionization and heat
calibration at low energy (0.1~keV$_{ee}$ at 10~keV$_{ee}$) correspond to
$\sim$~2~\% uncertainty on the efficiency in the 28.4~-~86.6~keV range.
The quadratic sum of these uncertainties is 4~\%.
This attests the simplicity and robustness of the data analysis
of the EDELWEISS heat-and-ionzation detectors.\\
A common systematic uncertainty to all bolometric experiments
is the determination of the quenching factor of the heat or phonon signal.
Present direct~\cite{bib-quench1} and indirect~\cite{bib-quench2}
measurements are compatible with unity at the $\sim$~10~\% level.
If taken as an uncertainty, this range correspond to a $\sim$~10~\% variation
of the limit for $M_W$~=~100~GeV/c$^2$, increasing up to $\sim$~20~\% at 50~GeV/c$^2$.\\
The present limits are very similar to our previously
published results (see Fig.~\ref{fig:fig-EDW-Limits}). These limits
can also be expressed simply in terms of rate of nuclear recoils between
30 and 100~keV, a range over which the detector efficiency is approximately
constant and equal to 90~\% in all configurations (see Fig.~\ref{fig:fig-efficmc}). In 2000-2002, no
events were observed in a fiducial exposure of 13.6~kg$\cdot$d. Taking into account 
the 90~\% efficiency for nuclear recoil over this energy range, this 
corresponds to an effective exposure of 12.2~kg$\cdot$d). It results in a 90~\% C.L. limit
of 0.19~events/kg/day for nuclear recoils between 30 and 100~keV.
A similar rate limit is derived from the 2003 data set alone: the
3 events observed between 30 and 100~keV in the effective exposure
of 43.5~kg$\cdot$d correspond to a limit of 0.15~events/kg/d at
90~\% C.L.. For the combined data set, the effective exposure is 
55.8~kg$\cdot$d
and the limit at 90~\% C.L. is 0.12~events/kg/d between 30 and 100~keV.\\
In Fig.~\ref{fig:fig-EDW-Limits}, the present limits
are compared with other experiments (CDMS~\cite{bib-cdms-suf,bib-cdms-soudan}
and CRESST~\cite{bib-cresst}). Because of the observed events, the EDELWEISS-I
limits are a factor 3 to 4 higher than the limits given by CDMS-II,
where surface events are efficiently rejected by phonon timing 
cuts~\cite{bib-cdms-soudan}.
\section{\label{sec:conclusion}Conclusion}
The EDELWEISS collaboration has searched for nuclear
recoils due to the scattering of WIMP dark matter using several 320~g
heat-and-ionization Ge detectors operated in a low-background environment
in the Laboratoire Souterrain de Modane. Up to three detectors have
been operated simultaneously, with consistent results. In the final
EDELWEISS-I setup, stable running conditions were achieved over periods
of four months with a recoil energy threshold better than 13~keV on
the three detectors. In the total fiducial exposure of 62~kg$\cdot$d,
40 nuclear recoil candidates are recorded between 15 and 200~keV.
Three of them are between 30 and 100~keV, a critical energy range
for establishing limits on WIMP interactions in the present experiment.
The study of detector coincidences and of the charge collection reveals
the presence of two likely sources of background: a residual neutron
background and surface electron-recoil events. Nevertheless, the limits
obtained on spin-independent WIMP-nucleon scattering cross-section
are very similar to the previously published results based on the
initial 13.6~kg$\cdot$d exposure~\cite{bib-edw2002}. The present results
supersede those of Refs.~\cite{bib-edw2000,bib-edw2002}.\\
The successful operation of the EDELWEISS-I setup
has provided important information for the preparation of the EDELWEISS-II
phase. The experimental volume in the EDELWEISS-I setup was limited to one liter. 
In the new setup under construction, the larger size dilution
cryostat (50~$\ell$) will be able to house up to 120 detectors, increasing the
rate at which exposure can be accumulated. The large number of detectors
(28 in a first stage) and the corresponding increase in coincidence
rate should facilitate the diagnostic regarding the actual level of
the residual neutron flux. This flux should also be drastically reduced
by the installation of a 50~cm polyethylene shielding offering a more
uniform coverage over all solid angles. In addition, a scintillating
muon veto surrounding the experiment should tag neutrons created by
muon interactions in the shielding. Regarding the surface electron
background, more care is taken in the selection of all materials surrounding
the detectors. The collaboration is also developing new detectors
with NbSi athermal phonon sensors that can tag surface 
events~\cite{bib-athermal-nbsi}.
\subsection*{Acknowledgments}
The help of the technical staff of the Laboratoire
Souterrain de Modane and the participant laboratories is gratefully
acknowledged. This work has been partially supported by the EEC Applied
Cryodetector network (Contracts ERBFMRXCT980167, HPRN-CT-2002-00322) and the ILIAS integrating
activity (Contract RII3-CT-2003-506222)


\newpage 

\newpage

\begin{table*}[tbp]
\caption{\label{tab:tab-Detectors}Main parameters of the detectors
used in EDELWEISS-I, i(p) denotes ionization (phonon) trigger conditions
(see Sec.~\ref{sec:acquisition}).}
\begin{ruledtabular}
\begin{tabular}{cccccccc}
Run&Detector&Mass (g)&\multicolumn{2}{c}{Amorphous layer}&
Al electrode &$T_{running}$ (mK)&$V_{bias}$ (V)\\
&&&Material&Thickness (nm)&thickness (nm)&&\\
\hline
2000&GeAl6&321.6&none&&100&27&+6.34\\
\hline
2002&GGA1&318.5&Ge&60&70&17&-4\\
\hline
2003i&GSA3&297.0&Si&25&64&17&-4\\
and&GSA1&313.7&Si&50&70&17&-4\\
2003p&GGA3&324.4&Ge&50&100&17&-4\\
\end{tabular}
\end{ruledtabular}
\end{table*}

\begin{table*}[tbp]
\caption{\label{tab:tab-Data-resos}Full-width at half-maximum
(FWHM) resolutions (in keV$_{ee}$) for heat and ionization signals
obtained for the detectors used in EDELWEISS-I, typical errors are
less than $\sim$~10~\%.}
\begin{ruledtabular}
\begin{tabular}{cccccccc}
Run&Detector&\multicolumn{4}{c}{Baseline resolution (keV$_{ee}$)}&
\multicolumn{2}{c}{Resolution at 122~keV$_{ee}$ (keV$_{ee}$)}\\
&&\multicolumn{3}{c}{Ionization}&Heat&Ionization&Heat\\
&&Center&Guard&Total&&&\\
\hline
2000&GeAl6&2.0&1.4&2.4&2.2&2.8&3.5\\
\hline
2002&GGA1&1.3&1.3&1.8&1.3&2.8&3.5\\
\hline
&GSA3&1.1&1.2&1.6&1.6&2.1&3.0\\
2003i&GSA1\footnote{ Quality 1 data (see Sec.~\ref{sec:data-new} for explanation).}&1.4&1.1&1.8&2.4&2.6&4.0\\
&GSA1\footnote{ Quality 2 data (see Sec.~\ref{sec:data-new} for explanation).}&1.4&1.5&2.1&4.6&2.8&5.0\\
&GGA3&1.7&2.0&2.6&0.44&3.5&3.2\\
\hline
&GSA3&1.1&1.4&1.8&1.7&2.4&3.0\\
2003p&GSA1&1.2&1.4&1.8&0.80&2.8&1.4\\
&GGA3&1.1&1.6&1.9&0.40&3.1&2.5\\
\end{tabular}
\end{ruledtabular}
\end{table*}

\begin{table*}[tbp]
\caption{\label{tab:tab-Data-sets}Thresholds for EDELWEISS-I low-background
run data sets. The quoted thresholds correspond to the energy at which
the efficiency for nuclear recoils reaches half its maximum value of 90~\%. 
The ionization
and heat thresholds are measured at trigger level 
(see Sec.~\ref{sec:calibration}), with uncertainties
less than $\pm$~0.5~keV$_{ee}$; the recoil energy thresholds are
measured after all analysis cuts, with a $\pm$~1~keV uncertainty.
The simulated recoil energy thresholds are obtained for 
$M_{W}$~=~100~GeV/c$^{2}$ (see Sec.~\ref{sec:simu})
with uncertainties less than $\pm$~1~keV.}
\begin{ruledtabular}
\begin{tabular}{ccccccc}
&&Ionization&Heat&Measured recoil&
Simulated recoil&Fiducial volume\\
Run&Detector&threshold&threshold&energy threshold&energy threshold&exposure\\
&&(keV$_{ee}$)&(keV$_{ee}$)&(keV)&(keV)&(kg$\cdot$d)\\
\hline
&&5.7&&&23&3.80\\
2000&GeAl6&9.0&&&31&0.63\\
&&11.0&&&37&0.60\\
\hline
2002&GGA1&3.5&&14&14&8.6\\
\hline
&GSA3&3.3&&14&13&9.16\\
2003i&GSA1\footnote{ Quality 1 data (see Sec.~\ref{sec:data-new} for explanation).}&4.6&&18&18&2.37\\
&GSA1\footnote{ Quality 2 data (see Sec.~\ref{sec:data-new} for explanation).}&4.6&&&24&2.81\\
&GGA3&5.8&&23&21&11.31\\
\hline
&GSA3&2.5&4.3&13&12&7.20\\
2003p&GSA1&2.5&2.3&12&11&7.60\\
&GGA3&2.5&1.6&11&11&7.86\\
\end{tabular}
\end{ruledtabular}
\end{table*}

\newpage

\end{document}